\documentclass[%
 reprint,
 amsmath,amssymb,
 aps,
pra,
]{revtex4-2}
\usepackage[utf8]{inputenc}
\usepackage[T1]{fontenc}
\usepackage{graphicx}
\usepackage{dcolumn}
\usepackage{bm}
\usepackage[utf8]{inputenc} 
\DeclareUnicodeCharacter{025B}{\ensuremath{\varepsilon}}
\usepackage{color,tcolorbox}

\usepackage[pdftex,colorlinks=true,linkcolor=Cerulean,citecolor=Cerulean,urlcolor=Cerulean]{hyperref}
\hypersetup{linkcolor=blue,citecolor=blue,urlcolor=blue}

\begin{document}

\preprint{APS/123-QED}

\title{Symmetries and conservation of spin angular momentum, helicity, and chirality in photonic time-varying media}

\author{Mohsen Mohammadi Jajin}
\email{mohsen.mohammadijajin@unavarra.es}
\affiliation{Department of Electrical, Electronic and Communications Engineering, Institute of Smart Cities (ISC), Universidad P\'ublica de Navarra (UPNA), 31006 Pamplona, Spain}
\author{J. Enrique V\'azquez-Lozano}
\email{enrique.vazquez@unavarra.es}
\affiliation{Department of Electrical, Electronic and Communications Engineering, Institute of Smart Cities (ISC), Universidad P\'ublica de Navarra (UPNA), 31006 Pamplona, Spain}
\author{I\~nigo Liberal}
\email{inigo.liberal@unavarra.es}
\affiliation{Department of Electrical, Electronic and Communications Engineering, Institute of Smart Cities (ISC), Universidad P\'ublica de Navarra (UPNA), 31006 Pamplona, Spain}

\begin{abstract}
Polarization-dependent dynamical properties of light as the spin angular momentum (SAM), helicity, and chirality are conserved quantities in free-space. Despite their similarities on account of their relationship with a circular state of polarization, SAM, helicity, and chirality emerge from distinct symmetries, which endows them with different physical meanings, properties, and practical applications. In this work, we investigate the behavior of such quantities in time-varying media (TVM), i.e., how a temporal modulation impacts their symmetries and conservation laws. Our results demonstrate that the SAM is conserved for any time modulation, helicity is only preserved in impedance-matched time modulations, while chirality is not conserved. In addition, the continuity equations highlight the dependence of the chirality with the energy content of the fields. These results provide additional insights into the similarities and differences between SAM, helicity, and chirality, as well as their physical meaning. Furthermore, our theoretical framework provides with a new perspective to analyze polarization-dependent light-matter interactions in TVM.
\end{abstract}

\maketitle

\section{Introduction}

By introducing time as an extra degree of freedom for taming light-matter interactions, time-varying media (TVM, also often referred to as temporal metamaterials) considerably stretch out the domains of optics and nanophotonics \cite{Engheta2021,Galiffi2022, Yin2022, Yuan2022, Engheta2023}. Indeed, the ability to actively and dynamically modulate optical material properties opens up new avenues for electromagnetic field manipulation. Recent examples of these possibilities include surpassing bandwidth bounds in impedance matching \cite{Shlivinski2018}, compact and low-energy nonreciprocal devices \cite{Sounas2017}, quantum state frequency shifting and ultra-fast switching without thermal noise amplification \cite{Liberal2023}, inverse prism and temporal aiming effects \cite{Akbarzadeh2018,Pacheco2020}, energy accumulation without a theoretical limit \cite{Mirmoosa2019} and control over the polarization of light \cite{galiffi2022archimedes,yin2022temporal,mostafa2024temporal}. Furthermore, TVM enable new mechanisms of photon generation and amplification \cite{Pendry2021}, including directional vacuum amplification effects \cite{VazquezLozano2023A}, enhanced light emission from quantum emitters \cite{Lyubaro2022}, and the design of incandescent sources that are not limited to the conventional blackbody spectrum \cite{VazquezLozano2023B}.

Noether's theorem states that any continuous symmetry has a corresponding conserved quantity \cite{KosmannSchwarzbach,CohenTannoudji,Sakurai,Fushchich,Banados2016,OrtegaGomez2023}. Therefore, much of the new physics provided by TVM emerge from the breaking of temporal symmetries. For example, breaking continuous translation symmetry removes the constraints associated with energy, allowing for light amplification, absorption, and emission \cite{Pendry2021,VazquezLozano2023A,Lyubaro2022,VazquezLozano2023B}. Similarly, the restrictions imposed by reciprocity are removed when temporal translation symmetry is broken \cite{Sounas2017}. Despite this fact, TVM can preserve certain spatial symmetries. For example, TVM with a continuous time-translation symmetry conserve the Minkowski momentum of the electromagnetic field \cite{OrtegaGomez2023}.

Interestingly, the electromagnetic field has a much wider range of symmetries and conserved quantities \cite{Kibble1965,Fushchich1992,Lipkin1964}. Hence, besides energy \cite{Loudon1970,Ruppin2002}, linear momentum \cite{Philbin2011,Bliokh2017}, and angular momentum (AM) \cite{Philbin2012,Bliokh2017}, including its separated components \cite{Birula2011}, i.e., the spin (SAM) and the orbital angular (OAM) momenta \cite{NietoVesperinas2015,Barnett2016}, the electromagnetic field exhibits  other polarization-dependent conserved quantities, such as helicity \cite{Cameron2012,FernandezCorbaton2013,Alpeggiani2018} and chirality \cite{Tang2010,Cameron2017,VazquezLozano2018}. Determining the form of the conservation laws through the corresponding continuity equations \cite{Nienhuis2016,FernandezCorbaton2017}, the associated symmetries, and the conditions under which they keep being fulfilled is of both fundamental and practical interest. However, unlike the former dynamical properties, the symmetries bringing forth the conservation of helicity and chirality are no pure spatial or temporal continuous transformations. Instead, they are mixed spatial and temporal transformations, also simultaneously involving electric and magnetic fields and/or potentials. In this sense, whether these symmetries and their associated conservation laws continue to be hold in TVM remains an open question.

We note that studying the particular cases of near-zero-index (NZI) \cite{lobet2022momentum} and time-varying \cite{OrtegaGomez2023} media was very illustrative in shedding some light onto the resolution of the Abraham-Minkowski debate \cite{Barnett2010A,Barnett2010B,Silveirinha2017}. Similarly, TVM is expected to further illustrate the current discussion on the differences between SAM, helicity, and chirality \cite{Philbin2013,Crimin2019,Mackinnon2019,Poulikakos2019}.

Following this motivation, in this work we investigate how temporal modulation affects polarization-dependent conserved quantities of the electromagnetic field. Specifically, building upon the framework of the Lagrangian theory of the electromagnetic fields, we look into the symmetries related to SAM, helicity, and chirality, assessing if they continue to be a symmetry of the electromagnetic field for TVM. Then, we determine the associated conserved quantity via Noether's theorem, showing the relationship with the conserved quantities in the non-time-modulated case. Finally, we investigate the continuity equations of SAM, helicity, and chirality, confirming the analysis based on the Lagrangian formalism, while obtaining insights on their potential sources and sinks under time modulation.

\begin{figure*}[t!]
\centering
\includegraphics[width=\textwidth,height=\textheight,keepaspectratio]{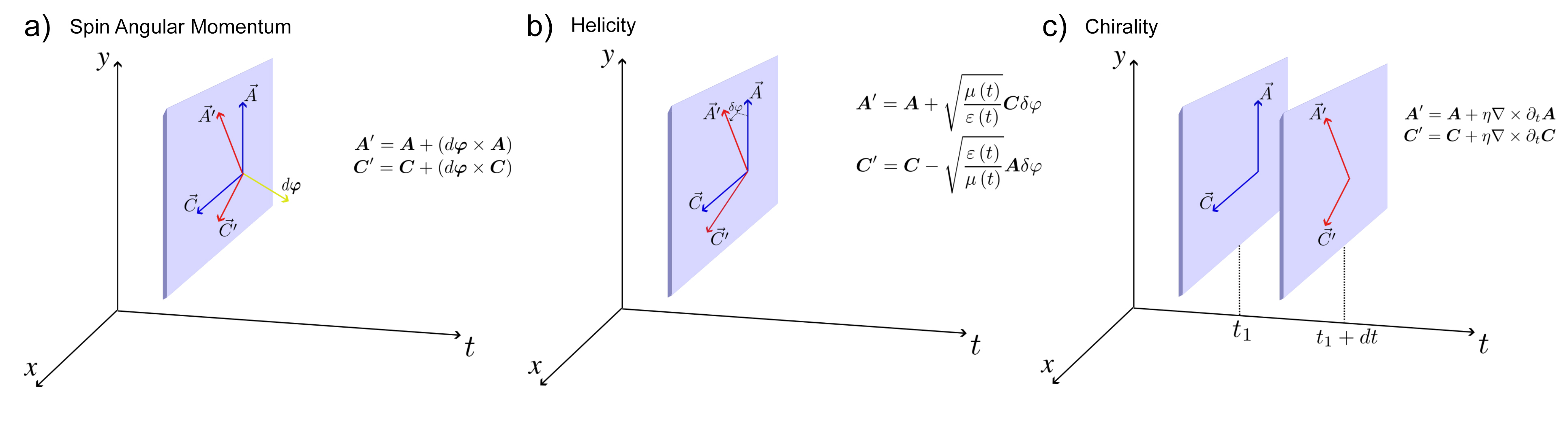}
\caption{{\bf Symmetries associated with polarization-dependent conserved quantities}. (a) The symmetry of the spin anfular momentum (SAM) represented as an infinitesimal rotation of the vector potentials about $\theta$. (b) Helicity is associated with an infinitesimal duality transformation exchanging electric and magnetic vector potentials. (c) The symmetry corresponding to the conservation of chirality consists of mixed of differential spatial and temporal transformations.}
\label{Figure}
\end{figure*}

\section{Dual-symmetric Lagrangian formalism in time-varying media}

\subsection{Lagrangian description of time-varying media}\label{lagrangian}

We assume an instantaneous and homogeneous time-varying medium (TVM), characterized in terms of time-dependent macroscopic constitutive parameters, namely, permittivity $\varepsilon(t)$, and permeability $\mu(t)$. For this configuration, the time-domain electric $\boldsymbol{E} \left(\mathbf{r},t\right)$ and magnetic $\boldsymbol{H}\left(\mathbf{r},t\right)$ fields are found as the solution to Maxwell curl equations
\begin{equation}
\label {Maxwell}
\begin{split}
& \nabla\times\boldsymbol{E}\left(\mathbf{r},t\right)=-\partial_{t}\left\{ \mu\left(t\right)\boldsymbol{H}\left(\mathbf{r},t\right)\right\} \\\\
&\nabla\times\boldsymbol{H}\left(\mathbf{r},t\right)=\partial_{t}\left\{ \varepsilon\left(t\right)\boldsymbol{E}\left(\mathbf{r},t\right)\right\} 
\end{split}
\end{equation}

A further extension of this model would include the use of more realistic material models including dispersion and loss ~\cite{hayran2022,Solis,gratus2021}. However, we note that the presence of dispersion and loss precludes the existence of any conserved quantity \cite{VazquezLozano2018,Alpeggiani2018,Bliokh2017,Philbin2011,Philbin2012}. In addition, there is a widespread use of instantaneous time-varying permittivity and permeability in the literature \cite{gratus2021,mai2023,zurita2009,gaxiola2021,ramaccia2020,ramaccia2021}.

Alternatively, the dynamics of the fields can be derived from a Lagrangian formalism \cite{CohenTannoudji}
\begin{equation}
L\left(t\right)=\int d^{3}\mathbf{r}\,\mathcal{L}\left(\mathbf{r},t\right)
\end{equation}

\noindent where the Lagrangian density for TVM is given by ~\cite{OrtegaGomez2023}
\begin{equation}
\label{Lagrang}
\mathcal{L}\left(\mathbf{r},t\right)=\frac{1}{2}\left[\varepsilon\left(t\right)\boldsymbol{E}\left(\mathbf{r},t\right)^2-\mu\left(t\right)\boldsymbol{H}\left(\mathbf{r},t\right)^2\right]
\end{equation}

The Lagrangian density given by Eq.\,(\ref{Lagrang}) is a direct extension from the non-time-modulated case, justified by the fact that the associated Lagrange’s equation recover Maxwell equations ~\cite{OrtegaGomez2023}.

Several works have pointed out that while Maxwell equations (\ref{Maxwell}) are dual symmetric, the non-time-modulated version of the Lagrangian density (\ref{Lagrang}) is not ~\citep{CohenTannoudji,kong1975}. The TVM version of the Lagrangian introduced in \cite{OrtegaGomez2023} suffers from the same shortcoming. This disagreement between the symmetries of Maxwell equations and the Lagrangian can be solved by reformulating the Lagrangian in terms of magnetic $\boldsymbol{A}\left(\mathbf{r},t\right)$ and electric $\boldsymbol{C}\left(\mathbf{r},t\right)$ vector potentials. Furthermore, we will choose a generalized Coulomb gauge as $\nabla\cdot\boldsymbol{A}\left(\mathbf{r},t\right)=0$, $\nabla\cdot\boldsymbol{C}\left(\mathbf{r},t\right)=0$, and consider scenarios in the absence of charges, so that all fields are transversal and can be obtained from the vector potentials as follows
\begin{equation}
\label{Vectors}
\begin{split}
&\boldsymbol{E}\left(\mathbf{r},t\right)=-\partial_{t}\boldsymbol{A}\left(\mathbf{r},t\right)=-\frac{1}{\varepsilon\left(t\right)}\nabla\times\boldsymbol{C}\left(\mathbf{r},t\right)\\
&\boldsymbol{H}\left(\mathbf{r},t\right)=-\partial_{t}\boldsymbol{C}\left(\mathbf{r},t\right)=\frac{1}{\mu\left(t\right)}\nabla\times\boldsymbol{A}\left(\mathbf{r},t\right)
\end{split}
\end{equation}

Putting these definitions into Maxwell's equations leads to the wave equations for the vector potentials,
\begin{equation}
\begin{split}
\label {Waveequationmaxwell}
&\nabla\times\nabla\times\boldsymbol{C}\left(\mathbf{r},t\right)=-\varepsilon\left(t\right)\partial_{t}\left\{ \mu\left(t\right)\partial_{t}\boldsymbol{C}\left(\mathbf{r},t\right)\right\}  \\
& \nabla\times\nabla\times\boldsymbol{A}\left(\mathbf{r},t\right)=-\mu\left(t\right)\partial_{t}\left\{ \varepsilon\left(t\right)\partial_{t}\mathbf{\boldsymbol{A}}\left(\mathbf{r},t\right)\right\}  
\end{split}
\end{equation}

\noindent which, on account of the Coulomb gauge, in component notation can be simply expressed as
\begin{equation}
\begin{split}
&\nabla^{2}A_{p}\left(\mathbf{r},t\right)=\mu\left(t\right)\partial_{t}\left\{ \varepsilon\left(t\right)\partial_{t}{A}_{p}\left(\mathbf{r},t\right)\right\}  \\
&\nabla^{2}C_{p}\left(\mathbf{r},t\right)=\varepsilon\left(t\right)\partial_{t}\left\{ \mu\left(t\right)\partial_{t}C_{p}\left(\mathbf{r},t\right)\right\}  
\end{split}
\end{equation}

\indent Then, we postulate the following a Lagrangian density for the electromagnetic field in TVM written in
terms of the vector potentials, $\boldsymbol{A}\left(\mathbf{r},t\right)$ and $\boldsymbol{C}\left(\mathbf{r},t\right)$
\begin{equation}
\label {Lagrangian}
\begin{split}
&\mathcal{L}\left(\mathbf{r},t\right)=\frac{1}{2}\left[\varepsilon\left(t\right)\left(\partial_{t}\boldsymbol{A}\right)^2-\frac{1}{\mu\left(t\right)}\left(\nabla\times\boldsymbol{A}\right)^2\right.\\
&+\left.\mu\left(t\right)\left(\partial_{t}\boldsymbol{C}\right)^2-\frac{1}{\varepsilon\left(t\right)}\left(\nabla\times\boldsymbol{C}\right)^2\right]
\end{split}
\end{equation}

\indent The Lagrangian density for TVM given by Eq.\,(\ref{Lagrangian}) is justified by the fact that Lagrange’s equations recover the wave equations for the vector potentials (\ref{Waveequationmaxwell}), while exhibiting the duality symmetry in the non-time-modulated limit \cite{bliokh2013dual}. Furthermore, it is worth highlighting the democratic form of the Lagrangian, featuring both electric and magnetic potentials on equal footing \cite{berry2009optical,avetisyan2021democratic}

\subsection{Noether's theorem, symmetries and conserved quantities}\label{Noether's theorem}

Noether's theorem states that each continuous symmetry of the Lagrangian is associated with a conserved quantity  \cite{OrtegaGomez2023,KosmannSchwarzbach,CohenTannoudji,Sakurai,Banados2016,Fushchich}. A continuous symmetry can be defined as a differential transformation of the dynamical variables that keeps the action’s integral unchanged, thus leading to the same equations of motion. Hence, for a Lagrangian density written in terms of the dynamical variables ${A}_{p}\left(\mathbf{r},t\right)$ and ${C}_{p}\left(\mathbf{r},t\right)$, a continuous symmetry is characterized by the differential variations $d{A}_{p}\left(\mathbf{r},t\right)={A}'_{p}\left(\mathbf{r},t\right)-{A}_{p}\left(\mathbf{r},t\right)$ and $d{C}_{p}\left(\mathbf{r},t\right)={C}'_{p}\left(\mathbf{r},t\right)-{C}_{p}\left(\mathbf{r},t\right)$. The associated variation of the action can be expressed as
\begin{equation}
dS=S'-S=\int_{t_{1}'}^{t_{2}'}dt\,L\left(A_{p}',C_{p}'\right)-\int_{t_{1}}^{t_{2}}dt\,L\left(A_{p},C_{p}\right) 
\end{equation}

\indent Integrating by parts and after some mathematical rearrangements involving the Euler-Lagrange equations, it can be demonstrated that a zero variation of the action, $dS=0$, requires the following quantity to be conserved in time
\begin{equation}
\label {Conservation}
\begin{split}
&\Psi\left(dA_{p},dC_{p}\right)=\sum_{p}\int d^{3}r\left\{ \frac{\partial L}{\partial\partial_{t}A_{p}\left(\mathbf{r},t\right)}dA_{p}\left(\mathbf{r},t\right)\right.\\
&\left.+\frac{\partial L}{\partial\partial_{t}C_{p}\left(\mathbf{r},t\right)}dC_{p}\left(\mathbf{r},t\right)\right\}   
\end{split}
\end{equation}

In this manner, Noether's theorem highlights that when the electromagnetic Lagrangian exhibits a continuous symmetry, described in terms of $dA_{p}$ and $dC_{p}$, the corresponding action must stay unchanged through an associated conserved quantity $\Psi\left(dA_{p},dC_{p}\right)$.

\section{Polarization-dependent symmetries and conserved quantitites in time-varying media}

Having established the dual-symmetric Lagrangian for TVM, and the form of Noether's theorem emerging from it, next we apply this theoretical framework to investigate how different polarization-dependent symmetries and conserved quantities are affected by the time-modulation of the material parameters. 

\subsection{Spin Angular Momentum (SAM)}\label{sec:SAM}

Spin angular momentum (SAM) is a quantity inherently connected to light polarization ~\cite{Birula2011,NietoVesperinas2015,Barnett2016,Cameron2012,bliokh2013dual}. Specifically, it explicitly refers to the circular polarization state, which, in the particular case of a plane wave, is aligned along the direction of propagation, determined either by the wavevector, ${\bf k}$, or the Poynting vector, ${\bf S}$.  Beyond describing the polarization of light, the SAM is at the basis of several fundamental optical phenomena, such as the spin-orbit interaction of light ~\cite{bliokh2015} , the spin Hall effect ~\cite{haefner2009,bliokh2015q}, or the spin-momentum locking ~\cite{van2016}.

Regarding its conserved character based on symmetry aspects, the SAM is related to infinitesimal rotations of the fields \cite{Cameron2012}. Specifically, the SAM is associated with the following differential transformations
\begin{equation}
\begin{split}
 & \boldsymbol{E}'=\boldsymbol{E}+\left(d\boldsymbol{\varphi}\times\boldsymbol{E}\right)\\
&\boldsymbol{B}'=\boldsymbol{B}+\left(d\boldsymbol{\varphi}\times\boldsymbol{B}\right)\\
\end{split}
\end{equation}

\noindent which can be equivalently written as a function of the vector potentials as follows
\begin{equation}
\begin{split}
 & \boldsymbol{A}'=\boldsymbol{A}+\left(d\boldsymbol{\varphi}\times\boldsymbol{A}\right)\\
&\boldsymbol{C}'=\boldsymbol{C}+\left(d\boldsymbol{\varphi}\times\boldsymbol{C}\right)\\
\end{split}
\end{equation}

Such symmetry is illustrated in Figure \ref{Figure}(a), which shows the variation of vector potentials under transformation of this symmetry in time-varying media. Writing the transformation  in component form 
\begin{equation}
\begin{split}
 & A'_{p}=A_{p}+\epsilon_{pij}d{\varphi}_{i}A_{j}\\
&C'_{p}=C_{p}+\epsilon_{pij}d{\varphi}_{i}C_{j}\\
\end{split}
\label{eq:sym_SAM_AC}
\end{equation}

Taking the Laplacian of Eq.\,(\ref{eq:sym_SAM_AC}) and rearranging the terms we find that the wave equations for the vector potentials are preserved. In other words, we can write
\begin{equation}
\begin{split}
 &\nabla^{2}A'_{p}=\mu\left(t\right)\partial_{t}\left\{ \varepsilon\left(t\right)\partial_{t}A'_{p}\right\} \\
&\nabla^{2}C'_{p}=\varepsilon\left(t\right)\partial_{t}\left\{ \mu\left(t\right)\partial_{t}C_{p}'\right\} \\
\end{split}
\end{equation}

Since the wave equations, i.e., the equations of motion, are preserved, it can be stated that the symmetry given  by Eq.\,(\ref{eq:sym_SAM_AC}) continues to be a symmetry even in the presence of time-modulation. This result could be expected since the symmetry described by Eq.\,(\ref{eq:sym_SAM_AC}) is purely a spatial symmetry. Therefore, modulating the constitutive parameters in time should not have an impact on such symmetry. Following Noether's theorem, such symmetry is associated with a conserved quantity. Specifically, by introducing (\ref{eq:sym_SAM_AC}) into (\ref{Conservation}), we find that such conserved quantity is
\[
\Psi=-\frac{1}{2}\int d^{3}r\left\{ \varepsilon\left(t\right)\boldsymbol{E}\cdot\left(d\boldsymbol{\varphi}\times\boldsymbol{A}\right)+\mu\left(t\right)\boldsymbol{H}\cdot\left(d\boldsymbol{\varphi}\times\boldsymbol{C}\right)\right\} 
\]
\begin{equation}
\label{Spina}
=\frac{1}{2}d\boldsymbol{\varphi}\cdot\int d^{3}r\left\{ \varepsilon\left(t\right)\left(\boldsymbol{E}\times\boldsymbol{A}\right)
+\mu\left(t\right)\left(\boldsymbol{C}\times\boldsymbol{H}\right)\right\} 
\end{equation}

According to the prescriptions of Noether's theorem, the above result shows the existence of a conserved quantity, the SAM of light, whose spatial density would be given by:
\begin{equation}
\mathbf{J}_{S}\left(\mathbf{r},t\right)=\frac{1}{2}\left\{ \varepsilon\left(t\right)\left(\boldsymbol{E}\times\boldsymbol{A}\right)+\mu\left(t\right)\left(\boldsymbol{H}\times\boldsymbol{C}\right)\right\} 
\label{eq:J_S}
\end{equation}

It can be concluded from Eq.\,(\ref{eq:J_S}) that the conserved quantity is a direct extension of the usual form of the SAM to TVM. Therefore, it is demonstrated that the SAM of light is a conserved quantity even in time-varying media. Again, this result stems from the fact that the SAM is rooted on a purely spatial symmetry. Following the same considerations, it should be expected that the orbital angular momentum (OAM) would be conserved too, and, accordingly, the total angular momentum (AM) should also be conserved. Our theory provides a rigorous theoretical framework to check this intuition, as it is properly justified in the Appendix. Besides its inherent theoretical interest, the conservation of the SAM also entails practical applications for extending spin-orbit interactions and the spin Hall effect of light to TVM.

\subsection{Helicity}\label{sec:Helicity}

Helicity is a fundamental property of spinning particles that, in the context of optics, is also related to the polarization of the electromagnetic field \citep{Alpeggiani2018,Cameron2012,FernandezCorbaton2013}. However, while both SAM and helicity are related to circular polarization, they capture different aspects of the wave's behavior~\cite{trueba,bliokh2013dual}. Indeed, while SAM is a vector quantity that represents the intrinsic angular momentum associated with the wave's polarization state, helicity is a scalar quantity often related to the alignment between the direction of propagation and the wave's angular momentum~\citep{Cameron2012}. Thus, the electromagnetic helicity is a well-defined quantity for the case of a circularly-polarized plane wave in free-space, though its range of applicability spans many other systems, including metamaterials and plasmonic structures ~\cite{negoro2023,matyushkin2020,olmos2022helicity}. As we will show, the existence of a temporal modulation accentuates the differences between SAM and helicity.

In the absence of time modulation, helicity is also a conserved quantity that emerges from the duality symmetry, corresponding to an exchange between electric and magnetic fields ~\cite{FernandezCorbaton2013}. In passing, we note that the duality symmetry in the antenna and microwave engineering community is often regarded as the duality theorem \cite{balanis}, and it is considered a convenient tool to reformulate and simplify electromagnetic problems. Specifically the duality symmetry can be expressed as follows ~\cite{Cameron2012,FernandezCorbaton2013}
\begin{equation}
\label{Duality}
\begin{split}
&\boldsymbol{E}'\left(\mathbf{r},t\right)\rightarrow\boldsymbol{E}\left(\mathbf{r},t\right)\mathrm{cos}\phi+Z\left(t\right)\boldsymbol{H}\left(\mathbf{r},t\right)\mathrm{sin}\phi\\
&\boldsymbol{H}'\left(\mathbf{r},t\right)\rightarrow-\frac{1}{Z\left(t\right)}\boldsymbol{E}\left(\mathbf{r},t\right)\mathrm{sin}\phi+\boldsymbol{H}\left(\mathbf{r},t\right)\mathrm{cos}\phi
\end{split}
\end{equation}

\noindent where we have defined the medium impedance $Z\left(t\right)=\sqrt{\mu\left(t\right)/\varepsilon\left(t\right)}$.

It is clear from Eq.\,(\ref{Duality}) that the duality symmetry is neither purely spatial nor temporal, as it intertwines electric and magnetic fields. As schematically depicted in Fig.\,\ref{Figure}(b), for a infinitesimal rotation of $\varphi$, and written in terms of vector potentials, the symmetry reduces to
\begin{equation}
\begin{split}
&\boldsymbol{A}'=\boldsymbol{A}+Z\left(t\right)\boldsymbol{C}\delta\varphi\\
&\boldsymbol{C}'=\boldsymbol{C}-\frac{1}{Z\left(t\right)}\boldsymbol{A}\delta\varphi
\end{split}
\label{eq:Sym_Helicity}
\end{equation}

Next, we double check if such (\ref{eq:Sym_Helicity}) continuous to be a symmetry of Maxwell equations in TVM. To this end we introduce (\ref{eq:Sym_Helicity}) into (\ref{Waveequationmaxwell}) leading to
\begin{equation}
\begin{split}
&\nabla\times\nabla\times\mathbf{A}'\left(\mathbf{r},t\right)=-\mu\left(t\right)\left[\partial_{t}\left\{ \varepsilon\left(t\right)\partial_{t}\mathbf{\mathbf{A}}'\left(\mathbf{r},t\right)\right\}\right. \\
&\left.+\delta\varphi\frac{1}{Z\left(t\right)}\partial_{t}\left\{ \mu\left(t\right)\partial_{t}\mathbf{C}'\left(\mathbf{r},t\right)\right\} \right]
\end{split}
\label{eq:WE_A_nex}
\end{equation}

It is then proven that the wave equations of the vector potentials are not preserved for arbitrary time modulations $\mu\left(t\right)$ and $\varepsilon\left(t\right)$. Consequently, it can be concluded that, in general, duality is not a symmetry in TVM. However, let us consider the particular case of TVM where the permittivity and permeability display the same time evolution, i.e.,
\begin{equation}
\label{Evolution}
\begin{split}
&\mu\left(t\right)=\mu\left(0\right)f\left(t\right)\\
&\varepsilon\left(t\right)=\varepsilon\left(0\right)f\left(t\right)
\end{split}
\end{equation}

For this particular case, the wave equation (\ref{eq:WE_A_nex}) recovers its original form. Intuitively, it can be understood that the time modulation described by Eq.\,(\ref {Evolution}) represents an impedance-matched modulation, i.e., $Z\left(t\right)=Z\left(0\right)$, which preserves the proportionality between electric and magnetic fields, thus preserving a duality symmetry.

Next, we confirm that the conserved quantity associated with such symmetry will indeed be the helicity. To this end, we substitute Eq.\,(\ref{eq:Sym_Helicity}) into the dual form of Lagrangian of the system, Eq.\,(\ref{Lagrangian}), it follows that the conserved quantity will be
\begin{equation}
\begin{split}
&\Psi\left(dA_{p},dC_{p}\right)=\frac{1}{2}\sum_{p}\int d^{3}r\left\{ \varepsilon\left(t\right)\partial_{t}A_{p}\left(Z\left(t\right)C_{p}\delta\varphi\right)\right.\\
&\left.+\mu\left(t\right)\partial_{t}C_{p}\left(-\frac{1}{Z\left(t\right)}A_{p}\delta\varphi\right)\right\} 
\end{split}
\end{equation}

Consequently, it is confirmed that duality symmetry leads to the conservation of helicity in the system, with helicity density
\begin{equation}
h\left(\mathbf{r},t\right)=\frac{1}{2}\left(\frac{1}{Z\left(t\right)}\boldsymbol{B}\cdot\boldsymbol{A} 
-Z\left(t\right)\boldsymbol{D}\cdot\boldsymbol{C}\right) 
\label{eq:Helicity}
\end{equation}

Hence, we have shown that the duality symmetry giving rise to the conservation of helicity is not generally preserved for TVM with arbitrary modulation profiles. Despite this fact, helicity is conserved in the particular case of impedance-matched temporal modulations. From these results, it is worth stressing out that, despite being polarization-dependent quantities, SAM and helicity exhibit different conservation laws and symmetries. The SAM has a purely spatial symmetry and it is therefore conserved for all temporal modulations. On the other hand, electromagnetic helicity has duality symmetry that intertwines electric and magnetic fields/potentials, which is not generally preserved in TVM.

\subsection{Chirality}\label{sec:Chirality}

Chirality is a geometrical property that allows us to describe objects that cannot be superimposed with their mirror image. This property is ubiquitous in nature, from physical objects to electromagnetic waves. Indeed, in the context of optics and nanophotonics, chirality was originally introduced by Tang and Cohen as a fundamental scalar property of light describing the local handedness, or knottedness, of optical fields in free space ~\cite{Tang2010}. Thus, circularly polarized light, which can be either right- or left-handed, depending on the sense of rotation of the fields along the wave propagation, is typically considered as the paradigmatic example of chiral light. Therefore, regardless of the direction of propagation of waves, optical chirality is a property of the polarization state of waves \cite{bliokh2011characterizing}, and, hence, it is closely related to the SAM and helicity ~\cite{Philbin2013,Crimin2019,Mackinnon2019,Poulikakos2019}. At the same time, chirality characterizes a different aspects of electromagnetic waves. Specifically, chirality is a scalar that describes the handedness of the polarization state of a classical electromagnetic wave~\cite{bliokh2011characterizing}.

Moreover, chirality has important implications in the study of chiroptical light-matter interactions, such as chiral molecules and materials, as it enables a direct characterization on the differential optical response of chiral matter when interacting either with right or left-handed circularly polarized light. Furthermore, optical chirality is of practical relevance to customary sensing and spectroscopic methods \citep{yoo}, such as circular dichroism \cite{formen2024,olmos2024capturing}, enantiomeric discrimination \cite{huang2024}, or chiral biological and chemical samples \cite{Hendry2010}.

From a fundamental perspective, chirality is a conserved dynamical property of light in the absence of time modulation \cite{Tang2010,bliokh2011characterizing}. As schematically depicted in Fig.\,\ref{Figure}(c), its associated symmetry, expressed in terms of vector potentials, is given by ~\cite{Philbin2013}
\begin{equation}
\begin{split}
&\boldsymbol{A}'\left(\mathbf{r},t\right)=\boldsymbol{A}\left(\mathbf{r},t\right)+\eta\nabla\times\partial_{t}\boldsymbol{A}\left(\mathbf{r},t\right)\\
&\boldsymbol{C}'\left(\mathbf{r},t\right)=\boldsymbol{C}\left(\mathbf{r},t\right)+\eta\nabla\times\partial_{t}\boldsymbol{C}\left(\mathbf{r},t\right)\\
\end{split}
\label{eq:Sym_Chirality}
\end{equation}

By taking the double curl of (\ref{eq:Sym_Chirality}), and rearranging the terms, it can be demonstrated that this continuous transformation does not preserve the form of the wave equation for the vector potentials:
\begin{equation}
\begin{split}
&\nabla\times\nabla\times\boldsymbol{A}'\left(\mathbf{r},t\right)=\\
&-\mu\left(t\right)\partial_{t}\left(\varepsilon\left(t\right)\mathbf\partial_{t}{\mathbf{A}}'\left(\mathbf{r},t\right)\right)-\mu\left(t\right)\eta\partial_{t}\left(\partial_{t}\varepsilon\left(t\right)\nabla\times\partial_{t}\mathbf{A}\left(\mathbf{r},t\right)\right)\\
&-\eta\partial_{t}\mu\left(t\right)\partial_{t}\left\{ \varepsilon\left(t\right)\nabla\times\mathbf\partial_{t}{\mathbf{A}\left(\mathbf{r},t\right)}\right\} \\
&\nabla\times\nabla\times\boldsymbol{C}'\left(\mathbf{r},t\right)=\\
&\varepsilon\left(t\right)\partial_{t}\left(\mu\left(t\right)\mathbf\partial_{t}{\mathbf{C}}'\left(\mathbf{r},t\right)\right)+\varepsilon\left(t\right)\eta\partial_{t}\left(\partial_{t}\mu\left(t\right)\nabla\times\partial_{t}\mathbf{C}\left(\mathbf{r},t\right)\right)\\
&+\eta\partial_{t}\varepsilon\left(t\right)\partial_{t}\left\{ \mu\left(t\right)\nabla\times\partial_{t}\mathbf{C}\left(\mathbf{r},t\right)\right\} 
\end{split}
\label{eq:Wave_AC_Chirality}
\end{equation}

It can be concluded from (\ref{eq:Wave_AC_Chirality}) that the terms breaking the symmetry of the wave equation are all related to the time-derivative of the material parameters 
$\partial_t\mu$ and $\partial_t\varepsilon$. Therefore, it is concluded that (\ref{eq:Sym_Chirality}) is a symmetry of the wave equations in the absence of time modulation, but it stops being a symmetry in TVM. Consequently, chirality is a conserved quantity in the absence of time modulation, but it is not conserved in TVM. 

The fact that chirality is the conserved quantity associated with the symmetry (\ref{eq:Sym_Chirality}) can be demonstrated by introducing it into (\ref{Conservation}), leading to
\begin{equation}
\begin{split}
&\Psi\left(d\boldsymbol{A},d\boldsymbol{C}\right)=\sum_{p}\frac{1}{2}\int d^{3}r\left\{ \varepsilon\left(t\right)\partial_{t}\boldsymbol{A}\left(\eta\nabla\times\partial_{t}\boldsymbol{A}\right)\right.\\
&\left.+\mu\left(t\right)\partial_{t}\boldsymbol{C}\left(\eta\nabla\times\partial_{t}\boldsymbol{C}\right)\right\}  
\end{split}
\end{equation}
\begin{equation}
=\eta\frac{1}{2}\int d^{3}r\left\{ \varepsilon\left(t\right)\boldsymbol{E}\cdot\left(\nabla\times\boldsymbol{E}\right)+\mu\left(t\right)\boldsymbol{H}\cdot\left(\nabla\times\boldsymbol{H}\right)\right\} 
\end{equation}

This motivates the definition of the chirality density 

\begin{equation}
\label{chiral}
c\left(\mathbf{r},t\right)=\frac{1}{2}\left[\varepsilon\left(t\right)\boldsymbol{E}\cdot\left(\nabla\times\boldsymbol{E}\right)
+\mu\left(t\right)\boldsymbol{H}\cdot\left(\nabla\times\boldsymbol{H}\right)\right]
\end{equation}

This quantity is a direct extension of the usual definition of chirality \cite{VazquezLozano2018}, which is found not to be preserved in TVM.

In summary, the conservation properties of SAM, helicity, and chirality are governed by their respective symmetries. SAM exhibits purely spatial symmetry, rendering it a conserved quantity even in the presence of time modulation. In contrast, helicity possesses a distinct symmetry, duality, making its conservation contingent upon specific conditions related to media parameters. Chirality, characterized by simultaneous spatial and temporal symmetry, is not conserved in TVM. Therefore, it is confirmed that TVM emphasize the physical differences between the conserved quantities associated with the polarization of the electromagnetic field.   

\section{Continuity equations}

The previous section introduced the Lagrangian analysis of the continuous symmetries and associated conservation laws of polarization-dependent conserved quantities. In this section, we carry out an independent analysis from the perspective of the continuity equations for those quantities. As we will show, finding the continuity equations offer an independent way to determine whether a quantity is conserved. In addition, continuity equations provide additional insight by providing information on the sources and sinks for the quantities that are not conserved. 

\subsection{Spin Angular Momentum}

Following Eq.\,(\ref{eq:J_S}) in section \ref{sec:SAM}, the density of SAM of light in TVM is defined as follows
\begin{equation}
\mathbf{J}_{S}\left(\mathbf{r},t\right)=\frac{1}{2}\left\{ \varepsilon\left(t\right)\left(\boldsymbol{E}\times\boldsymbol{A}\right)+\mu\left(t\right)\left(\boldsymbol{H}\times\boldsymbol{C}\right)\right\} 
\end{equation}

By taking the time derivative, using Maxwell’s equations and the vector calculus identities, $\left(\nabla\times\boldsymbol{A}\right)\times\boldsymbol{B}\text{=}\nabla\left(\boldsymbol{B}\cdot\boldsymbol{A}\right)-\left(\boldsymbol{B}\cdot\nabla\right)\boldsymbol{A}-\left(\boldsymbol{A}\cdot\nabla\right)\boldsymbol{B}$, we can compactly write the continuity equation
\begin{equation}
\partial_{t}\boldsymbol{S} + \nabla\cdot\overline{\overline{\boldsymbol{F}}}_{S}=0
\label{eq:Continuity_SAM}
\end{equation}

\noindent where we have introduced the flux density of SAM as
\begin{equation}
\begin{split}
&\overline{\overline{\boldsymbol{F}}}_{S}=-\frac{1}{2}\left[\overline{\overline{I}}\left(\boldsymbol{A}\cdot\boldsymbol{H}\right)-\boldsymbol{A}\boldsymbol{H}-\boldsymbol{H}\boldsymbol{A}\right.\\
&\left.-\overline{\overline{I}}\left(\boldsymbol{C}\cdot\boldsymbol{E}\right)+\boldsymbol{C}\boldsymbol{E}+\boldsymbol{E}\boldsymbol{C}\right]
\end{split}
\end{equation}

\noindent where $\overline{\overline{I}}$ is the identity dyadic, and $\boldsymbol{A}\boldsymbol{H}$ is the dyadic product of $\boldsymbol{A}$ and $\boldsymbol{H}$.

It can be concluded from Eq.\,(\ref{eq:Continuity_SAM}) that the continuity equation of the SAM in TVM does not involve source/sink terms on its right hand side. Therefore, it is independently derived that the SAM is a conserved quantity in TVM. 

\subsection{Helicity}

Similarly, the helicity density is defined in accordance with Eq.\,(\ref{eq:Helicity}) in Section \ref{sec:Helicity}: 
\begin{equation}
h\left(\mathbf{r},t\right)=\frac{1}{2}\left(\frac{1}{Z\left(t\right)}\boldsymbol{B}\cdot\boldsymbol{A} 
-Z\left(t\right)\boldsymbol{D}\cdot\boldsymbol{C}\right) 
\label{ref:Helicity_2}
\end{equation}

Taking the temporal derivative of (\ref{ref:Helicity_2}), using Maxwell's equations, and the vector calculus identity $\nabla\cdot\left(\boldsymbol{A}\times\boldsymbol{B}\right)=\boldsymbol{B}\cdot\left(\nabla\times\boldsymbol{A}\right)-\boldsymbol{A}\cdot\left(\nabla\times\boldsymbol{B}\right)$, we find the following continuity equation for the Helicity in TVM:
\begin{equation}
\begin{split}
&\frac{dh\left(\mathbf{r},t\right)}{dt}+\nabla\cdot\boldsymbol{F}_{h}\left(\mathbf{r},t\right)=\\
&\left[\frac{\mu\left(t\right)\partial_{t}\varepsilon\left(t\right)}{4\sqrt{\varepsilon\left(t\right)\mu\left(t\right)}}-\frac{\varepsilon\left(t\right)\partial_{t}\mu\left(t\right)}{4\sqrt{\varepsilon\left(t\right)\mu\left(t\right)}}\right]
\left[\boldsymbol{A}\cdot\boldsymbol{H}+\boldsymbol{C}\cdot\boldsymbol{E}\right]\\
\end{split}
\label{eq:Helicity_continuity}
\end{equation}

\noindent where we have defined the helicity flux density as
\begin{equation}
\boldsymbol{F}_{h}\left(\mathbf{r},t\right)=
\frac{1}{2}\left[
\frac{1}{Z\left(t\right)}\left(\boldsymbol{E}\times\boldsymbol{A}\right)
+Z\left(t\right)\left(\boldsymbol{H}\times\boldsymbol{C}\right)
\right]
\end{equation}

It is confirmed in Eq.\,(\ref{eq:Helicity_continuity}) that helicity is not a conserved quantity in TVM. In particular, its continuity equation presents source/sink terms directly related to the time derivatives of the permittivity and permeability. At the same time, it can be double-checked that such source/sink terms vanish for the specific case of an impedanced-matched time modulation: 
$\varepsilon\left(t\right)=\varepsilon\left(0\right)f\left(t\right)$ and $ \mu\left(t\right)=\mu\left(0\right)f\left(t\right)$. Therefore, the continuity equation confirms the conclusion that helicity is conserved in TVM only for media that preserves the medium impedance, i.e., maintaining the symmetry between electric and magnetic fields. 

\subsection{Chirality}

Finally, following Eq.\,(\ref{chiral}) in Section \ref{sec:Chirality}, the chirality density is defined as follows
\begin{equation}
c\left(\mathbf{r},t\right)=
\frac{1}{2}\left[\varepsilon\left(t\right)\boldsymbol{E}\cdot\nabla\times\boldsymbol{E}
+\mu\left(t\right)\boldsymbol{H}\cdot\nabla\times\boldsymbol{H}\right]
\label{eq:Chirality_2}
\end{equation}

By taking the time derivative of (\ref{eq:Chirality_2}), and rearranging the terms, it is possible to find the following continuity equation for the chirality in TVM:
\[
\frac{dc\left(\mathbf{r},t\right)}{dt}+\nabla\cdot\boldsymbol{F}_c\left(\mathbf{r},t\right)=
\]
\begin{equation}
-\frac{1}{2}\left[\partial_{t}\varepsilon\boldsymbol{E}\cdot\left(\nabla\times\boldsymbol{E}\right)+\partial_{t}\mu\boldsymbol{H}\cdot\left(\nabla\times\boldsymbol{H}\right)\right]
\label{eq:Chirality_continuity}
\end{equation}

\noindent where we have defined the chirality flux density as
\begin{equation}
\boldsymbol{F}_c=\frac{1}{2}\left[\boldsymbol{E}\times\left(\nabla\times\boldsymbol{H}\right)-\boldsymbol{H}\times\left(\nabla\times\boldsymbol{E}\right)\right]
\end{equation}

Similar to helicity, it is found that the continuity equation for the chirality in TVM presents source/sink terms related to the time-derivatives of the permittivity and permeability. Therefore, it is also confirmed that chirality is not a conserved quantity in TVM. In this case, it is not straightforward to identify a specific temporal modulation that would lead to the conservation of chirality. 
However, some particular cases present an interesting behavior. For example, let us assume an impedance-matched temporal modulation preserving helicy, i.e.,  
$\varepsilon\left(t\right)=\varepsilon\left(0\right)f\left(t\right)$ and $ \mu\left(t\right)=\mu\left(0\right)f\left(t\right)$. In this case, the continuity equation for the optical chirality reduces to

\begin{equation}
\frac{dc\left(\mathbf{r},t\right)}{dt}+\nabla\cdot\boldsymbol{F}_{c}\left(\mathbf{r},t\right)
=-\frac{1}{f\left(t\right)}\,\frac{df\left(t\right)}{dt}\,c\left(\mathbf{r},t\right)
\end{equation}

Then, by considering the total chirality of the system

\begin{equation}
C\left(t\right)=\int d^{3}r\,c\left(\mathbf{r},t\right)
\end{equation}

\noindent and assuming that all fields are contained within the region of integration, we find that it obeys the following differential equation
\begin{equation}
\frac{dC\left(t\right)}{dt}=-\frac{1}{f\left(t\right)}\,\frac{df\left(t\right)}{dt}\,C\left(t\right)
\end{equation}

\noindent with the solution
\begin{equation}
C\left(t\right)=C\left(0\right)\mathrm{exp}\left\{-\mathrm{ln}\left(f\left(t\right)\right)\right\} 
\end{equation}

It is worth pointing out that, if we carry out the same calculation for the energy of the system
\begin{equation}
\mathcal{H}\left(t\right)=\frac{1}{2}\int d^{3}r\,\left[\varepsilon\left(t\right)\boldsymbol{E}^{2}\left(\mathbf{r},t\right)+\mu\left(t\right)\boldsymbol{H}^{2}\left(\mathbf{r},t\right)\right],
\end{equation}

\noindent we find that the time-derivative of the energy is given by 
\begin{equation}
\frac{d\mathcal{H}\left(t\right)}{dt}=-\left[\frac{d\varepsilon\left(t\right)}{dt}\boldsymbol{E}^{2}\left(\mathbf{r},t\right)+\frac{d\mu\left(t\right)}{dt}\boldsymbol{H}^{2}\left(\mathbf{r},t\right)\right]
\end{equation}

Thus, for the particular case of an impedanced-matched temporal modulation the solution would be given by 
\begin{equation}
\mathcal{H}\left(t\right)=
\mathcal{H}\left(0\right)\mathrm{exp}\left\{-\mathrm{ln}\left(f\left(t\right)\right)\right\} 
\end{equation}

Therefore, it is found that while chirality is not conserved in an impedance-matched time-modulated medium, it directly scales with the energy of the electromagnetic field. In fact, several works have highlighted that an impedance-matched time modulation results in a frequency shift of the fields, thus a change in energy, while preserving the overall field distribution ~\cite{Pendry2021,Liberal2023}. Therefore, this analysis suggests that helicity is more dependent on the polarization of the electromagnetic field, irrespectively of the energy content, while chirality is more directly linked to the energy content of the fields.  

\section{Conclusion}

Our results clarify the symmetries and conservation of SAM, helicity and chirality in TVM. The different conservation rules can be intuitively understood in accordance to the spatial, temporal, or mixed nature of their associated symmetries. At the same time, they are derived from a rigorous formalism that highlights the need of a dual-symmetric Lagrangian to correctly account for polarization-dependent quantities in time-varying media. We found that the SAM is conserved in TVM, which might be of interest for generalizations of the spin-orbit interaction of light, the spin Hall effect or the spin-momentum locking to TVM, as well as applications related to optical forces. On the contrary, we find that chirality and helicity are in general not conserved in TVM. This result points towards future research on using TVM in engineering chirality in view of their applications for circular dichroism, enantiomeric discrimination, or chiral biological and chemical samples.

\begin{acknowledgments}
\textbf{Funding.} H2020 European Research Council (Starting Grant 948504); Ministerio de Ciencia, Innovación y Universidades
(MCIU) (FJC2021-047776-I, RYC2018-024123-I).

\textbf{Disclosures.} The authors declare no conflicts of interest.
\end{acknowledgments}

\section*{Appendix}\label{Appendix}

The main text has focused on symmetries and conserved quantities related to the polarization of the electromagnetic field. However, there are many more symmetries and conserved quantities. In this Appendix, we analyze the orbital angular momentum (OAM), and total angular momentum (AM). Similar to the SAM, these conserved quantities emerge from purely spatial symmetries and, therefore, they are also conserved in spatially homogeneous time-varying media.

\subsection*{Orbital Angular Momentum}

Another property that results from the phase distribution of a light wave's spatial distribution is the orbital angular momentum (OAM). Just like the SAM is associated with polarization, the OAM is related to the twisted or helical nature of certain light beams \citep{Barnett2016}. Applications of this property are found in optical communication, where information can be duplicated using various OAM states, thereby increasing the capacity of optical communication systems. The variation of the field in a rotation $d\varphi$ about the axis  is ~\citep{CohenTannoudji}
\begin{equation}
\begin{array}{c}
\boldsymbol{A}'=\boldsymbol{A}+\left[\left(d\boldsymbol{\varphi}\times\mathbf{r}\right)\cdot\nabla\right]\boldsymbol{A}\\
\boldsymbol{C}'=\boldsymbol{C}+\left[\left(d\boldsymbol{\varphi}\times\mathbf{r}\right)\cdot\nabla\right]\boldsymbol{C}
\end{array}
\end{equation}

\noindent where it is clear that the symmetry associated with the OAM is a pure spatial transformation. Similar to the SAM, it is possible to take the double curl of the symmetry and demonstrate that the wave equation of the potential preserves its original form. Therefore, the OAM is a conserved quantity even in the presence of time modulation. The associated conserved quantity is given by
\[
\Psi\left(dA_{p},dC_{p}\right)=d\boldsymbol{\varphi}\cdot\sum_{p}\int d^{3}r\left\{ \varepsilon\left(t\right)\partial_{t}A_{p}\left(\left(\mathbf{r}\times\nabla\right)A_{p}\right)\right.
\]
\begin{equation}
\left.-\mu\left(t\right)\partial_{t}C_{p}\left(\left(\mathbf{r}\times\nabla\right)C_{p}\right)\right\} 
\end{equation}

It is then clear that the conserved quantity is indeed the extension of the OAM to TVM, with density
\begin{equation}
\label{Orbital}
\mathbf{J}_{O}=\varepsilon\left(t\right)\boldsymbol{E}\cdot\left(\mathbf{r}\times\nabla\right)\boldsymbol{A}
+\mu\left(t\right)\boldsymbol{H}\left(\mathbf{r}\times\nabla\right)\boldsymbol{C}
\end{equation}

\subsection*{Total Angular Momentum}

Having demonstrated the conservation of the SAM and the OAM, it is clear that the total angular momentum (AM) should also be conserved. This fact can be demonstrated by defining the symmetry of the AM as the addition of both symmetries
\begin{equation}
\begin{array}{c}
d\boldsymbol{A}_{AM}'=d\boldsymbol{A}_{SAM}'+d\boldsymbol{A}_{OAM}'\\
d\boldsymbol{C}_{AM}'=d\boldsymbol{C}_{SAM}'+d\boldsymbol{C}_{OAM}'
\end{array}
\end{equation}

\noindent and then noting that the variation of the action (\ref{Conservation}) is also linear with the continuous transformations $d\boldsymbol{A}$ and $d\boldsymbol{C}$. Consequently, the conserved quantity will be the angular momentum, and it will have a density given by the addition of the SAM and OAM densities. 

\bibliography{library}

\providecommand{\noopsort}[1]{}\providecommand{\singleletter}[1]{#1}%
\begin{thebibliography}{78}%
\makeatletter
\providecommand \@ifxundefined [1]{%
 \@ifx{#1\undefined}
}%
\providecommand \@ifnum [1]{%
 \ifnum #1\expandafter \@firstoftwo
 \else \expandafter \@secondoftwo
 \fi
}%
\providecommand \@ifx [1]{%
 \ifx #1\expandafter \@firstoftwo
 \else \expandafter \@secondoftwo
 \fi
}%
\providecommand \natexlab [1]{#1}%
\providecommand \enquote  [1]{``#1''}%
\providecommand \bibnamefont  [1]{#1}%
\providecommand \bibfnamefont [1]{#1}%
\providecommand \citenamefont [1]{#1}%
\providecommand \href@noop [0]{\@secondoftwo}%
\providecommand \href [0]{\begingroup \@sanitize@url \@href}%
\providecommand \@href[1]{\@@startlink{#1}\@@href}%
\providecommand \@@href[1]{\endgroup#1\@@endlink}%
\providecommand \@sanitize@url [0]{\catcode `\\12\catcode `\$12\catcode
  `\&12\catcode `\#12\catcode `\^12\catcode `\_12\catcode `\%12\relax}%
\providecommand \@@startlink[1]{}%
\providecommand \@@endlink[0]{}%
\providecommand \url  [0]{\begingroup\@sanitize@url \@url }%
\providecommand \@url [1]{\endgroup\@href {#1}{\urlprefix }}%
\providecommand \urlprefix  [0]{URL }%
\providecommand \Eprint [0]{\href }%
\providecommand \doibase [0]{https://doi.org/}%
\providecommand \selectlanguage [0]{\@gobble}%
\providecommand \bibinfo  [0]{\@secondoftwo}%
\providecommand \bibfield  [0]{\@secondoftwo}%
\providecommand \translation [1]{[#1]}%
\providecommand \BibitemOpen [0]{}%
\providecommand \bibitemStop [0]{}%
\providecommand \bibitemNoStop [0]{.\EOS\space}%
\providecommand \EOS [0]{\spacefactor3000\relax}%
\providecommand \BibitemShut  [1]{\csname bibitem#1\endcsname}%
\let\auto@bib@innerbib\@empty
\bibitem [{\citenamefont {Engheta}(2021)}]{Engheta2021}%
  \BibitemOpen
  \bibfield  {author} {\bibinfo {author} {\bibfnamefont {N.}~\bibnamefont
  {Engheta}},\ }\bibfield  {title} {\bibinfo {title} {Metamaterials with high
  degrees of freedom: space, time, and more},\ }\href
  {https://doi.org/doi:10.1515/nanoph-2020-0414} {\bibfield  {journal}
  {\bibinfo  {journal} {Nanophotonics}\ }\textbf {\bibinfo {volume} {10}},\
  \bibinfo {pages} {639} (\bibinfo {year} {2021})}\BibitemShut {NoStop}%
\bibitem [{\citenamefont {Galiffi}\ \emph
  {et~al.}(2022{\natexlab{a}})\citenamefont {Galiffi}, \citenamefont {Tirole},
  \citenamefont {Yin}, \citenamefont {Li}, \citenamefont {Vezzoli},
  \citenamefont {Huidobro}, \citenamefont {Silveirinha}, \citenamefont
  {Sapienza}, \citenamefont {Al{\`u}},\ and\ \citenamefont
  {Pendry}}]{Galiffi2022}%
  \BibitemOpen
  \bibfield  {author} {\bibinfo {author} {\bibfnamefont {E.}~\bibnamefont
  {Galiffi}}, \bibinfo {author} {\bibfnamefont {R.}~\bibnamefont {Tirole}},
  \bibinfo {author} {\bibfnamefont {S.}~\bibnamefont {Yin}}, \bibinfo {author}
  {\bibfnamefont {H.}~\bibnamefont {Li}}, \bibinfo {author} {\bibfnamefont
  {S.}~\bibnamefont {Vezzoli}}, \bibinfo {author} {\bibfnamefont {P.~A.}\
  \bibnamefont {Huidobro}}, \bibinfo {author} {\bibfnamefont {M.~G.}\
  \bibnamefont {Silveirinha}}, \bibinfo {author} {\bibfnamefont
  {R.}~\bibnamefont {Sapienza}}, \bibinfo {author} {\bibfnamefont
  {A.}~\bibnamefont {Al{\`u}}},\ and\ \bibinfo {author} {\bibfnamefont {J.~B.}\
  \bibnamefont {Pendry}},\ }\bibfield  {title} {\bibinfo {title} {{Photonics of
  time-varying media}},\ }\href {https://doi.org/10.1117/1.AP.4.1.014002}
  {\bibfield  {journal} {\bibinfo  {journal} {Advanced Photonics}\ }\textbf
  {\bibinfo {volume} {4}},\ \bibinfo {pages} {014002} (\bibinfo {year}
  {2022}{\natexlab{a}})}\BibitemShut {NoStop}%
\bibitem [{\citenamefont {Yin}\ \emph {et~al.}(2022{\natexlab{a}})\citenamefont
  {Yin}, \citenamefont {Galiffi},\ and\ \citenamefont {Al{\`u}}}]{Yin2022}%
  \BibitemOpen
  \bibfield  {author} {\bibinfo {author} {\bibfnamefont {S.}~\bibnamefont
  {Yin}}, \bibinfo {author} {\bibfnamefont {E.}~\bibnamefont {Galiffi}},\ and\
  \bibinfo {author} {\bibfnamefont {A.}~\bibnamefont {Al{\`u}}},\ }\bibfield
  {title} {\bibinfo {title} {Floquet metamaterials},\ }\href
  {https://doi.org/10.1186/s43593-022-00015-1} {\bibfield  {journal} {\bibinfo
  {journal} {ELight}\ }\textbf {\bibinfo {volume} {2}},\ \bibinfo {pages} {1}
  (\bibinfo {year} {2022}{\natexlab{a}})}\BibitemShut {NoStop}%
\bibitem [{\citenamefont {Yuan}\ and\ \citenamefont {Fan}(2022)}]{Yuan2022}%
  \BibitemOpen
  \bibfield  {author} {\bibinfo {author} {\bibfnamefont {L.}~\bibnamefont
  {Yuan}}\ and\ \bibinfo {author} {\bibfnamefont {S.}~\bibnamefont {Fan}},\
  }\bibfield  {title} {\bibinfo {title} {Temporal modulation brings
  metamaterials into new era},\ }\href
  {https://doi.org/10.1038/s41377-022-00870-0} {\bibfield  {journal} {\bibinfo
  {journal} {Light: Science \& Applications}\ }\textbf {\bibinfo {volume}
  {11}},\ \bibinfo {pages} {173} (\bibinfo {year} {2022})}\BibitemShut
  {NoStop}%
\bibitem [{\citenamefont {Engheta}(2023)}]{Engheta2023}%
  \BibitemOpen
  \bibfield  {author} {\bibinfo {author} {\bibfnamefont {N.}~\bibnamefont
  {Engheta}},\ }\bibfield  {title} {\bibinfo {title} {Four-dimensional optics
  using time-varying metamaterials},\ }\href
  {https://doi.org/10.1126/science.adf1094} {\bibfield  {journal} {\bibinfo
  {journal} {Science}\ }\textbf {\bibinfo {volume} {379}},\ \bibinfo {pages}
  {1190} (\bibinfo {year} {2023})}\BibitemShut {NoStop}%
\bibitem [{\citenamefont {Shlivinski}\ and\ \citenamefont
  {Hadad}(2018)}]{Shlivinski2018}%
  \BibitemOpen
  \bibfield  {author} {\bibinfo {author} {\bibfnamefont {A.}~\bibnamefont
  {Shlivinski}}\ and\ \bibinfo {author} {\bibfnamefont {Y.}~\bibnamefont
  {Hadad}},\ }\bibfield  {title} {\bibinfo {title} {Beyond the bode-fano bound:
  Wideband impedance matching for short pulses using temporal switching of
  transmission-line parameters},\ }\href
  {https://doi.org/10.1103/PhysRevLett.121.204301} {\bibfield  {journal}
  {\bibinfo  {journal} {Physical review letters}\ }\textbf {\bibinfo {volume}
  {121}},\ \bibinfo {pages} {204301} (\bibinfo {year} {2018})}\BibitemShut
  {NoStop}%
\bibitem [{\citenamefont {Sounas}\ and\ \citenamefont
  {Al{\`u}}(2017)}]{Sounas2017}%
  \BibitemOpen
  \bibfield  {author} {\bibinfo {author} {\bibfnamefont {D.~L.}\ \bibnamefont
  {Sounas}}\ and\ \bibinfo {author} {\bibfnamefont {A.}~\bibnamefont
  {Al{\`u}}},\ }\bibfield  {title} {\bibinfo {title} {Non-reciprocal photonics
  based on time modulation},\ }\href
  {https://doi.org/10.1038/s41566-017-0051-x} {\bibfield  {journal} {\bibinfo
  {journal} {Nature Photonics}\ }\textbf {\bibinfo {volume} {11}},\ \bibinfo
  {pages} {774} (\bibinfo {year} {2017})}\BibitemShut {NoStop}%
\bibitem [{\citenamefont {Liberal}\ \emph {et~al.}(2023)\citenamefont
  {Liberal}, \citenamefont {V{\'a}zquez-Lozano},\ and\ \citenamefont
  {Pacheco-Pe{\~n}a}}]{Liberal2023}%
  \BibitemOpen
  \bibfield  {author} {\bibinfo {author} {\bibfnamefont {I.}~\bibnamefont
  {Liberal}}, \bibinfo {author} {\bibfnamefont {J.~E.}\ \bibnamefont
  {V{\'a}zquez-Lozano}},\ and\ \bibinfo {author} {\bibfnamefont
  {V.}~\bibnamefont {Pacheco-Pe{\~n}a}},\ }\bibfield  {title} {\bibinfo {title}
  {Quantum antireflection temporal coatings: quantum state frequency shifting
  and inhibited thermal noise amplification},\ }\href
  {https://doi.org/10.1002/lpor.202200720} {\bibfield  {journal} {\bibinfo
  {journal} {Laser \& Photonics Reviews}\ }\textbf {\bibinfo {volume} {17}},\
  \bibinfo {pages} {2200720} (\bibinfo {year} {2023})}\BibitemShut {NoStop}%
\bibitem [{\citenamefont {Akbarzadeh}\ \emph {et~al.}(2018)\citenamefont
  {Akbarzadeh}, \citenamefont {Chamanara},\ and\ \citenamefont
  {Caloz}}]{Akbarzadeh2018}%
  \BibitemOpen
  \bibfield  {author} {\bibinfo {author} {\bibfnamefont {A.}~\bibnamefont
  {Akbarzadeh}}, \bibinfo {author} {\bibfnamefont {N.}~\bibnamefont
  {Chamanara}},\ and\ \bibinfo {author} {\bibfnamefont {C.}~\bibnamefont
  {Caloz}},\ }\bibfield  {title} {\bibinfo {title} {Inverse prism based on
  temporal discontinuity and spatial dispersion},\ }\href
  {https://doi.org/10.1364/OL.43.003297} {\bibfield  {journal} {\bibinfo
  {journal} {Optics letters}\ }\textbf {\bibinfo {volume} {43}},\ \bibinfo
  {pages} {3297} (\bibinfo {year} {2018})}\BibitemShut {NoStop}%
\bibitem [{\citenamefont {Pacheco-Pe{\~n}a}\ and\ \citenamefont
  {Engheta}(2020)}]{Pacheco2020}%
  \BibitemOpen
  \bibfield  {author} {\bibinfo {author} {\bibfnamefont {V.}~\bibnamefont
  {Pacheco-Pe{\~n}a}}\ and\ \bibinfo {author} {\bibfnamefont {N.}~\bibnamefont
  {Engheta}},\ }\bibfield  {title} {\bibinfo {title} {Temporal aiming},\ }\href
  {https://doi.org/10.1038/s41377-020-00360-1} {\bibfield  {journal} {\bibinfo
  {journal} {Light: Science \& Applications}\ }\textbf {\bibinfo {volume}
  {9}},\ \bibinfo {pages} {129} (\bibinfo {year} {2020})}\BibitemShut {NoStop}%
\bibitem [{\citenamefont {Mirmoosa}\ \emph {et~al.}(2019)\citenamefont
  {Mirmoosa}, \citenamefont {Ptitcyn}, \citenamefont {Asadchy},\ and\
  \citenamefont {Tretyakov}}]{Mirmoosa2019}%
  \BibitemOpen
  \bibfield  {author} {\bibinfo {author} {\bibfnamefont {M.~S.}\ \bibnamefont
  {Mirmoosa}}, \bibinfo {author} {\bibfnamefont {G.}~\bibnamefont {Ptitcyn}},
  \bibinfo {author} {\bibfnamefont {V.~S.}\ \bibnamefont {Asadchy}},\ and\
  \bibinfo {author} {\bibfnamefont {S.~A.}\ \bibnamefont {Tretyakov}},\
  }\bibfield  {title} {\bibinfo {title} {Time-varying reactive elements for
  extreme accumulation of electromagnetic energy},\ }\href
  {https://doi.org/10.1103/PhysRevApplied.11.014024} {\bibfield  {journal}
  {\bibinfo  {journal} {Physical Review Applied}\ }\textbf {\bibinfo {volume}
  {11}},\ \bibinfo {pages} {014024} (\bibinfo {year} {2019})}\BibitemShut
  {NoStop}%
\bibitem [{\citenamefont {Galiffi}\ \emph
  {et~al.}(2022{\natexlab{b}})\citenamefont {Galiffi}, \citenamefont
  {Huidobro},\ and\ \citenamefont {Pendry}}]{galiffi2022archimedes}%
  \BibitemOpen
  \bibfield  {author} {\bibinfo {author} {\bibfnamefont {E.}~\bibnamefont
  {Galiffi}}, \bibinfo {author} {\bibfnamefont {P.~A.}\ \bibnamefont
  {Huidobro}},\ and\ \bibinfo {author} {\bibfnamefont {J.}~\bibnamefont
  {Pendry}},\ }\bibfield  {title} {\bibinfo {title} {An archimedes' screw for
  light},\ }\href {https://doi.org/10.1038/s41467-022-30079-z} {\bibfield
  {journal} {\bibinfo  {journal} {Nature Communications}\ }\textbf {\bibinfo
  {volume} {13}},\ \bibinfo {pages} {2523} (\bibinfo {year}
  {2022}{\natexlab{b}})}\BibitemShut {NoStop}%
\bibitem [{\citenamefont {Yin}\ \emph {et~al.}(2022{\natexlab{b}})\citenamefont
  {Yin}, \citenamefont {Wang},\ and\ \citenamefont
  {Al{\`u}}}]{yin2022temporal}%
  \BibitemOpen
  \bibfield  {author} {\bibinfo {author} {\bibfnamefont {S.}~\bibnamefont
  {Yin}}, \bibinfo {author} {\bibfnamefont {Y.-T.}\ \bibnamefont {Wang}},\ and\
  \bibinfo {author} {\bibfnamefont {A.}~\bibnamefont {Al{\`u}}},\ }\bibfield
  {title} {\bibinfo {title} {Temporal optical activity and chiral
  time-interfaces},\ }\href {https://doi.org/10.1364/OE.480199} {\bibfield
  {journal} {\bibinfo  {journal} {Optics Express}\ }\textbf {\bibinfo {volume}
  {30}},\ \bibinfo {pages} {47933} (\bibinfo {year}
  {2022}{\natexlab{b}})}\BibitemShut {NoStop}%
\bibitem [{\citenamefont {Mostafa}\ \emph {et~al.}(2024)\citenamefont
  {Mostafa}, \citenamefont {Mirmoosa}, \citenamefont {Sidorenko}, \citenamefont
  {Asadchy},\ and\ \citenamefont {Tretyakov}}]{mostafa2024temporal}%
  \BibitemOpen
  \bibfield  {author} {\bibinfo {author} {\bibfnamefont {M.}~\bibnamefont
  {Mostafa}}, \bibinfo {author} {\bibfnamefont {M.}~\bibnamefont {Mirmoosa}},
  \bibinfo {author} {\bibfnamefont {M.}~\bibnamefont {Sidorenko}}, \bibinfo
  {author} {\bibfnamefont {V.}~\bibnamefont {Asadchy}},\ and\ \bibinfo {author}
  {\bibfnamefont {S.}~\bibnamefont {Tretyakov}},\ }\bibfield  {title} {\bibinfo
  {title} {Temporal interfaces in complex electromagnetic materials: an
  overview},\ }\href {https://doi.org/10.1364/OME.516179} {\bibfield  {journal}
  {\bibinfo  {journal} {Optical Materials Express}\ }\textbf {\bibinfo {volume}
  {14}},\ \bibinfo {pages} {1103} (\bibinfo {year} {2024})}\BibitemShut
  {NoStop}%
\bibitem [{\citenamefont {Pendry}\ \emph {et~al.}(2021)\citenamefont {Pendry},
  \citenamefont {Galiffi},\ and\ \citenamefont {Huidobro}}]{Pendry2021}%
  \BibitemOpen
  \bibfield  {author} {\bibinfo {author} {\bibfnamefont {J.}~\bibnamefont
  {Pendry}}, \bibinfo {author} {\bibfnamefont {E.}~\bibnamefont {Galiffi}},\
  and\ \bibinfo {author} {\bibfnamefont {P.}~\bibnamefont {Huidobro}},\
  }\bibfield  {title} {\bibinfo {title} {Gain in time-dependent media—a new
  mechanism},\ }\href {https://doi.org/10.1364/JOSAB.427682} {\bibfield
  {journal} {\bibinfo  {journal} {JOSA B}\ }\textbf {\bibinfo {volume} {38}},\
  \bibinfo {pages} {3360} (\bibinfo {year} {2021})}\BibitemShut {NoStop}%
\bibitem [{\citenamefont {V{\'a}zquez-Lozano}\ and\ \citenamefont
  {Liberal}(2022)}]{VazquezLozano2023A}%
  \BibitemOpen
  \bibfield  {author} {\bibinfo {author} {\bibfnamefont {J.~E.}\ \bibnamefont
  {V{\'a}zquez-Lozano}}\ and\ \bibinfo {author} {\bibfnamefont
  {I.}~\bibnamefont {Liberal}},\ }\bibfield  {title} {\bibinfo {title} {Shaping
  the quantum vacuum with anisotropic temporal boundaries},\ }\href
  {https://doi.org/10.1515/nanoph-2022-0491} {\bibfield  {journal} {\bibinfo
  {journal} {Nanophotonics}\ }\textbf {\bibinfo {volume} {12}},\ \bibinfo
  {pages} {539} (\bibinfo {year} {2022})}\BibitemShut {NoStop}%
\bibitem [{\citenamefont {Lyubarov}\ \emph {et~al.}(2022)\citenamefont
  {Lyubarov}, \citenamefont {Lumer}, \citenamefont {Dikopoltsev}, \citenamefont
  {Lustig}, \citenamefont {Sharabi},\ and\ \citenamefont
  {Segev}}]{Lyubaro2022}%
  \BibitemOpen
  \bibfield  {author} {\bibinfo {author} {\bibfnamefont {M.}~\bibnamefont
  {Lyubarov}}, \bibinfo {author} {\bibfnamefont {Y.}~\bibnamefont {Lumer}},
  \bibinfo {author} {\bibfnamefont {A.}~\bibnamefont {Dikopoltsev}}, \bibinfo
  {author} {\bibfnamefont {E.}~\bibnamefont {Lustig}}, \bibinfo {author}
  {\bibfnamefont {Y.}~\bibnamefont {Sharabi}},\ and\ \bibinfo {author}
  {\bibfnamefont {M.}~\bibnamefont {Segev}},\ }\bibfield  {title} {\bibinfo
  {title} {Amplified emission and lasing in photonic time crystals},\ }\href
  {https://doi.org/10.1126/science.abo3324} {\bibfield  {journal} {\bibinfo
  {journal} {Science}\ }\textbf {\bibinfo {volume} {377}},\ \bibinfo {pages}
  {425} (\bibinfo {year} {2022})}\BibitemShut {NoStop}%
\bibitem [{\citenamefont {V{\'a}zquez-Lozano}\ and\ \citenamefont
  {Liberal}(2023)}]{VazquezLozano2023B}%
  \BibitemOpen
  \bibfield  {author} {\bibinfo {author} {\bibfnamefont {J.~E.}\ \bibnamefont
  {V{\'a}zquez-Lozano}}\ and\ \bibinfo {author} {\bibfnamefont
  {I.}~\bibnamefont {Liberal}},\ }\bibfield  {title} {\bibinfo {title}
  {Incandescent temporal metamaterials},\ }\href
  {https://doi.org/10.1038/s41467-023-40281-2} {\bibfield  {journal} {\bibinfo
  {journal} {Nature Communications}\ }\textbf {\bibinfo {volume} {14}},\
  \bibinfo {pages} {4606} (\bibinfo {year} {2023})}\BibitemShut {NoStop}%
\bibitem [{\citenamefont {Kosmann-Schwarzbach}\ \emph
  {et~al.}(2011)\citenamefont {Kosmann-Schwarzbach}, \citenamefont
  {Schwarzbach},\ and\ \citenamefont
  {Kosmann-Schwarzbach}}]{KosmannSchwarzbach}%
  \BibitemOpen
  \bibfield  {author} {\bibinfo {author} {\bibfnamefont {Y.}~\bibnamefont
  {Kosmann-Schwarzbach}}, \bibinfo {author} {\bibfnamefont {B.~E.}\
  \bibnamefont {Schwarzbach}},\ and\ \bibinfo {author} {\bibfnamefont
  {Y.}~\bibnamefont {Kosmann-Schwarzbach}},\ }\href@noop {} {\emph {\bibinfo
  {title} {The Noether Theorems}}}\ (\bibinfo  {publisher} {Springer},\
  \bibinfo {year} {2011})\BibitemShut {NoStop}%
\bibitem [{\citenamefont {Cohen-Tannoudji}\ \emph {et~al.}(1997)\citenamefont
  {Cohen-Tannoudji}, \citenamefont {Dupont-Roc},\ and\ \citenamefont
  {Grynberg}}]{CohenTannoudji}%
  \BibitemOpen
  \bibfield  {author} {\bibinfo {author} {\bibfnamefont {C.}~\bibnamefont
  {Cohen-Tannoudji}}, \bibinfo {author} {\bibfnamefont {J.}~\bibnamefont
  {Dupont-Roc}},\ and\ \bibinfo {author} {\bibfnamefont {G.}~\bibnamefont
  {Grynberg}},\ }\href@noop {} {\emph {\bibinfo {title} {Photons and
  atoms-introduction to quantum electrodynamics}}}\ (\bibinfo {year}
  {1997})\BibitemShut {NoStop}%
\bibitem [{\citenamefont {Sakurai}\ and\ \citenamefont
  {Commins}(1995)}]{Sakurai}%
  \BibitemOpen
  \bibfield  {author} {\bibinfo {author} {\bibfnamefont {J.~J.}\ \bibnamefont
  {Sakurai}}\ and\ \bibinfo {author} {\bibfnamefont {E.~D.}\ \bibnamefont
  {Commins}},\ }\href@noop {} {\bibinfo {title} {Modern quantum mechanics,
  revised edition}} (\bibinfo {year} {1995})\BibitemShut {NoStop}%
\bibitem [{\citenamefont {Fushchich}\ and\ \citenamefont
  {Nikitin}()}]{Fushchich}%
  \BibitemOpen
  \bibfield  {author} {\bibinfo {author} {\bibfnamefont {W.~I.}\ \bibnamefont
  {Fushchich}}\ and\ \bibinfo {author} {\bibfnamefont {A.}~\bibnamefont
  {Nikitin}},\ }\href@noop {} {\emph {\bibinfo {title} {Symmetries of
  Maxwell’s equations}}}\BibitemShut {NoStop}%
\bibitem [{\citenamefont {Banados}\ and\ \citenamefont
  {Reyes}(2016)}]{Banados2016}%
  \BibitemOpen
  \bibfield  {author} {\bibinfo {author} {\bibfnamefont {M.}~\bibnamefont
  {Banados}}\ and\ \bibinfo {author} {\bibfnamefont {I.}~\bibnamefont
  {Reyes}},\ }\bibfield  {title} {\bibinfo {title} {A short review on
  noether’s theorems, gauge symmetries and boundary terms},\ }\href
  {https://doi.org/10.1142/S0218271816300214} {\bibfield  {journal} {\bibinfo
  {journal} {International Journal of Modern Physics D}\ }\textbf {\bibinfo
  {volume} {25}},\ \bibinfo {pages} {1630021} (\bibinfo {year}
  {2016})}\BibitemShut {NoStop}%
\bibitem [{\citenamefont {Ortega-Gomez}\ \emph {et~al.}(2023)\citenamefont
  {Ortega-Gomez}, \citenamefont {Lobet}, \citenamefont {V{\'a}zquez-Lozano},\
  and\ \citenamefont {Liberal}}]{OrtegaGomez2023}%
  \BibitemOpen
  \bibfield  {author} {\bibinfo {author} {\bibfnamefont {A.}~\bibnamefont
  {Ortega-Gomez}}, \bibinfo {author} {\bibfnamefont {M.}~\bibnamefont {Lobet}},
  \bibinfo {author} {\bibfnamefont {J.~E.}\ \bibnamefont
  {V{\'a}zquez-Lozano}},\ and\ \bibinfo {author} {\bibfnamefont
  {I.}~\bibnamefont {Liberal}},\ }\bibfield  {title} {\bibinfo {title}
  {Tutorial on the conservation of momentum in photonic time-varying media},\
  }\href {https://doi.org/10.1364/OME.485540} {\bibfield  {journal} {\bibinfo
  {journal} {Optical Materials Express}\ }\textbf {\bibinfo {volume} {13}},\
  \bibinfo {pages} {1598} (\bibinfo {year} {2023})}\BibitemShut {NoStop}%
\bibitem [{\citenamefont {Kibble}(1965)}]{Kibble1965}%
  \BibitemOpen
  \bibfield  {author} {\bibinfo {author} {\bibfnamefont {T.}~\bibnamefont
  {Kibble}},\ }\bibfield  {title} {\bibinfo {title} {Conservation laws for free
  fields},\ }\href {https://doi.org/10.1063/1.1704363} {\bibfield  {journal}
  {\bibinfo  {journal} {Journal of Mathematical Physics}\ }\textbf {\bibinfo
  {volume} {6}},\ \bibinfo {pages} {1022} (\bibinfo {year} {1965})}\BibitemShut
  {NoStop}%
\bibitem [{\citenamefont {Fushchich}\ and\ \citenamefont
  {Nikitin}(1992)}]{Fushchich1992}%
  \BibitemOpen
  \bibfield  {author} {\bibinfo {author} {\bibfnamefont {W.}~\bibnamefont
  {Fushchich}}\ and\ \bibinfo {author} {\bibfnamefont {A.}~\bibnamefont
  {Nikitin}},\ }\bibfield  {title} {\bibinfo {title} {The complete sets of
  conservation laws for the electromagnetic field},\ }\href
  {https://doi.org/10.1088/0305-4470/25/5/004} {\bibfield  {journal} {\bibinfo
  {journal} {J. Phys. A: Math. Gen}\ }\textbf {\bibinfo {volume} {25}},\
  \bibinfo {pages} {L231} (\bibinfo {year} {1992})}\BibitemShut {NoStop}%
\bibitem [{\citenamefont {Lipkin}(1964)}]{Lipkin1964}%
  \BibitemOpen
  \bibfield  {author} {\bibinfo {author} {\bibfnamefont {D.~M.}\ \bibnamefont
  {Lipkin}},\ }\bibfield  {title} {\bibinfo {title} {Existence of a new
  conservation law in electromagnetic theory},\ }\href
  {https://doi.org/10.1063/1.1704165} {\bibfield  {journal} {\bibinfo
  {journal} {Journal of Mathematical Physics}\ }\textbf {\bibinfo {volume}
  {5}},\ \bibinfo {pages} {696} (\bibinfo {year} {1964})}\BibitemShut {NoStop}%
\bibitem [{\citenamefont {Loudon}(1970)}]{Loudon1970}%
  \BibitemOpen
  \bibfield  {author} {\bibinfo {author} {\bibfnamefont {R.}~\bibnamefont
  {Loudon}},\ }\bibfield  {title} {\bibinfo {title} {The propagation of
  electromagnetic energy through an absorbing dielectric},\ }\href
  {https://doi.org/10.1088/0305-4470/3/3/008} {\bibfield  {journal} {\bibinfo
  {journal} {Journal of Physics A: General Physics}\ }\textbf {\bibinfo
  {volume} {3}},\ \bibinfo {pages} {233} (\bibinfo {year} {1970})}\BibitemShut
  {NoStop}%
\bibitem [{\citenamefont {Ruppin}(2002)}]{Ruppin2002}%
  \BibitemOpen
  \bibfield  {author} {\bibinfo {author} {\bibfnamefont {R.}~\bibnamefont
  {Ruppin}},\ }\bibfield  {title} {\bibinfo {title} {Electromagnetic energy
  density in a dispersive and absorptive material},\ }\href
  {https://doi.org/10.1016/S0375-9601(01)00838-6} {\bibfield  {journal}
  {\bibinfo  {journal} {Physics letters A}\ }\textbf {\bibinfo {volume}
  {299}},\ \bibinfo {pages} {309} (\bibinfo {year} {2002})}\BibitemShut
  {NoStop}%
\bibitem [{\citenamefont {Philbin}(2011)}]{Philbin2011}%
  \BibitemOpen
  \bibfield  {author} {\bibinfo {author} {\bibfnamefont {T.~G.}\ \bibnamefont
  {Philbin}},\ }\bibfield  {title} {\bibinfo {title} {Electromagnetic energy
  momentum in dispersive media},\ }\href
  {https://doi.org/10.1103/PhysRevA.83.013823} {\bibfield  {journal} {\bibinfo
  {journal} {Physical Review A}\ }\textbf {\bibinfo {volume} {83}},\ \bibinfo
  {pages} {013823} (\bibinfo {year} {2011})}\BibitemShut {NoStop}%
\bibitem [{\citenamefont {Bliokh}\ \emph {et~al.}(2017)\citenamefont {Bliokh},
  \citenamefont {Bekshaev},\ and\ \citenamefont {Nori}}]{Bliokh2017}%
  \BibitemOpen
  \bibfield  {author} {\bibinfo {author} {\bibfnamefont {K.~Y.}\ \bibnamefont
  {Bliokh}}, \bibinfo {author} {\bibfnamefont {A.~Y.}\ \bibnamefont
  {Bekshaev}},\ and\ \bibinfo {author} {\bibfnamefont {F.}~\bibnamefont
  {Nori}},\ }\bibfield  {title} {\bibinfo {title} {Optical momentum, spin, and
  angular momentum in dispersive media},\ }\href
  {https://doi.org/10.1103/PhysRevLett.119.073901} {\bibfield  {journal}
  {\bibinfo  {journal} {Physical review letters}\ }\textbf {\bibinfo {volume}
  {119}},\ \bibinfo {pages} {073901} (\bibinfo {year} {2017})}\BibitemShut
  {NoStop}%
\bibitem [{\citenamefont {Philbin}\ and\ \citenamefont
  {Allanson}(2012)}]{Philbin2012}%
  \BibitemOpen
  \bibfield  {author} {\bibinfo {author} {\bibfnamefont {T.~G.}\ \bibnamefont
  {Philbin}}\ and\ \bibinfo {author} {\bibfnamefont {O.}~\bibnamefont
  {Allanson}},\ }\bibfield  {title} {\bibinfo {title} {Optical angular momentum
  in dispersive media},\ }\href {https://doi.org/10.1103/PhysRevA.86.055802}
  {\bibfield  {journal} {\bibinfo  {journal} {Physical Review A}\ }\textbf
  {\bibinfo {volume} {86}},\ \bibinfo {pages} {055802} (\bibinfo {year}
  {2012})}\BibitemShut {NoStop}%
\bibitem [{\citenamefont {Bialynicki-Birula}\ and\ \citenamefont
  {Bialynicka-Birula}(2011)}]{Birula2011}%
  \BibitemOpen
  \bibfield  {author} {\bibinfo {author} {\bibfnamefont {I.}~\bibnamefont
  {Bialynicki-Birula}}\ and\ \bibinfo {author} {\bibfnamefont {Z.}~\bibnamefont
  {Bialynicka-Birula}},\ }\bibfield  {title} {\bibinfo {title} {Canonical
  separation of angular momentum of light into its orbital and spin parts},\
  }\href {https://doi.org/10.1088/2040-8978/13/6/064014} {\bibfield  {journal}
  {\bibinfo  {journal} {Journal of Optics}\ }\textbf {\bibinfo {volume} {13}},\
  \bibinfo {pages} {064014} (\bibinfo {year} {2011})}\BibitemShut {NoStop}%
\bibitem [{\citenamefont {Nieto-Vesperinas}(2015)}]{NietoVesperinas2015}%
  \BibitemOpen
  \bibfield  {author} {\bibinfo {author} {\bibfnamefont {M.}~\bibnamefont
  {Nieto-Vesperinas}},\ }\bibfield  {title} {\bibinfo {title} {Optical torque:
  electromagnetic spin and orbital-angular-momentum conservation laws and their
  significance},\ }\href {https://doi.org/10.1103/PhysRevA.92.043843}
  {\bibfield  {journal} {\bibinfo  {journal} {Physical Review A}\ }\textbf
  {\bibinfo {volume} {92}},\ \bibinfo {pages} {043843} (\bibinfo {year}
  {2015})}\BibitemShut {NoStop}%
\bibitem [{\citenamefont {Barnett}\ \emph {et~al.}(2016)\citenamefont
  {Barnett}, \citenamefont {Allen}, \citenamefont {Cameron}, \citenamefont
  {Gilson}, \citenamefont {Padgett}, \citenamefont {Speirits},\ and\
  \citenamefont {Yao}}]{Barnett2016}%
  \BibitemOpen
  \bibfield  {author} {\bibinfo {author} {\bibfnamefont {S.~M.}\ \bibnamefont
  {Barnett}}, \bibinfo {author} {\bibfnamefont {L.}~\bibnamefont {Allen}},
  \bibinfo {author} {\bibfnamefont {R.~P.}\ \bibnamefont {Cameron}}, \bibinfo
  {author} {\bibfnamefont {C.~R.}\ \bibnamefont {Gilson}}, \bibinfo {author}
  {\bibfnamefont {M.~J.}\ \bibnamefont {Padgett}}, \bibinfo {author}
  {\bibfnamefont {F.~C.}\ \bibnamefont {Speirits}},\ and\ \bibinfo {author}
  {\bibfnamefont {A.~M.}\ \bibnamefont {Yao}},\ }\bibfield  {title} {\bibinfo
  {title} {On the natures of the spin and orbital parts of optical angular
  momentum},\ }\href {https://doi.org/10.1088/2040-8978/18/6/064004} {\bibfield
   {journal} {\bibinfo  {journal} {Journal of Optics}\ }\textbf {\bibinfo
  {volume} {18}},\ \bibinfo {pages} {064004} (\bibinfo {year}
  {2016})}\BibitemShut {NoStop}%
\bibitem [{\citenamefont {Cameron}\ \emph {et~al.}(2012)\citenamefont
  {Cameron}, \citenamefont {Barnett},\ and\ \citenamefont {Yao}}]{Cameron2012}%
  \BibitemOpen
  \bibfield  {author} {\bibinfo {author} {\bibfnamefont {R.~P.}\ \bibnamefont
  {Cameron}}, \bibinfo {author} {\bibfnamefont {S.~M.}\ \bibnamefont
  {Barnett}},\ and\ \bibinfo {author} {\bibfnamefont {A.~M.}\ \bibnamefont
  {Yao}},\ }\bibfield  {title} {\bibinfo {title} {Optical helicity, optical
  spin and related quantities in electromagnetic theory},\ }\href
  {https://doi.org/10.1088/1367-2630/14/5/053050} {\bibfield  {journal}
  {\bibinfo  {journal} {New Journal of Physics}\ }\textbf {\bibinfo {volume}
  {14}},\ \bibinfo {pages} {053050} (\bibinfo {year} {2012})}\BibitemShut
  {NoStop}%
\bibitem [{\citenamefont {Fernandez-Corbaton}\ \emph
  {et~al.}(2013)\citenamefont {Fernandez-Corbaton}, \citenamefont
  {Zambrana-Puyalto}, \citenamefont {Tischler}, \citenamefont {Vidal},
  \citenamefont {Juan},\ and\ \citenamefont
  {Molina-Terriza}}]{FernandezCorbaton2013}%
  \BibitemOpen
  \bibfield  {author} {\bibinfo {author} {\bibfnamefont {I.}~\bibnamefont
  {Fernandez-Corbaton}}, \bibinfo {author} {\bibfnamefont {X.}~\bibnamefont
  {Zambrana-Puyalto}}, \bibinfo {author} {\bibfnamefont {N.}~\bibnamefont
  {Tischler}}, \bibinfo {author} {\bibfnamefont {X.}~\bibnamefont {Vidal}},
  \bibinfo {author} {\bibfnamefont {M.~L.}\ \bibnamefont {Juan}},\ and\
  \bibinfo {author} {\bibfnamefont {G.}~\bibnamefont {Molina-Terriza}},\
  }\bibfield  {title} {\bibinfo {title} {Electromagnetic duality symmetry and
  helicity conservation for the macroscopic maxwell’s equations},\ }\href
  {https://doi.org/10.1103/PhysRevLett.111.060401} {\bibfield  {journal}
  {\bibinfo  {journal} {Physical review letters}\ }\textbf {\bibinfo {volume}
  {111}},\ \bibinfo {pages} {060401} (\bibinfo {year} {2013})}\BibitemShut
  {NoStop}%
\bibitem [{\citenamefont {Alpeggiani}\ \emph {et~al.}(2018)\citenamefont
  {Alpeggiani}, \citenamefont {Bliokh}, \citenamefont {Nori},\ and\
  \citenamefont {Kuipers}}]{Alpeggiani2018}%
  \BibitemOpen
  \bibfield  {author} {\bibinfo {author} {\bibfnamefont {F.}~\bibnamefont
  {Alpeggiani}}, \bibinfo {author} {\bibfnamefont {K.}~\bibnamefont {Bliokh}},
  \bibinfo {author} {\bibfnamefont {F.}~\bibnamefont {Nori}},\ and\ \bibinfo
  {author} {\bibfnamefont {L.}~\bibnamefont {Kuipers}},\ }\bibfield  {title}
  {\bibinfo {title} {Electromagnetic helicity in complex media},\ }\href
  {https://doi.org/10.1103/PhysRevLett.120.243605} {\bibfield  {journal}
  {\bibinfo  {journal} {Physical review letters}\ }\textbf {\bibinfo {volume}
  {120}},\ \bibinfo {pages} {243605} (\bibinfo {year} {2018})}\BibitemShut
  {NoStop}%
\bibitem [{\citenamefont {Tang}\ and\ \citenamefont {Cohen}(2010)}]{Tang2010}%
  \BibitemOpen
  \bibfield  {author} {\bibinfo {author} {\bibfnamefont {Y.}~\bibnamefont
  {Tang}}\ and\ \bibinfo {author} {\bibfnamefont {A.~E.}\ \bibnamefont
  {Cohen}},\ }\bibfield  {title} {\bibinfo {title} {Optical chirality and its
  interaction with matter},\ }\href
  {https://doi.org/10.1103/PhysRevLett.104.163901} {\bibfield  {journal}
  {\bibinfo  {journal} {Physical review letters}\ }\textbf {\bibinfo {volume}
  {104}},\ \bibinfo {pages} {163901} (\bibinfo {year} {2010})}\BibitemShut
  {NoStop}%
\bibitem [{\citenamefont {Cameron}\ \emph {et~al.}(2017)\citenamefont
  {Cameron}, \citenamefont {G{\"o}tte}, \citenamefont {Barnett},\ and\
  \citenamefont {Yao}}]{Cameron2017}%
  \BibitemOpen
  \bibfield  {author} {\bibinfo {author} {\bibfnamefont {R.~P.}\ \bibnamefont
  {Cameron}}, \bibinfo {author} {\bibfnamefont {J.~B.}\ \bibnamefont
  {G{\"o}tte}}, \bibinfo {author} {\bibfnamefont {S.~M.}\ \bibnamefont
  {Barnett}},\ and\ \bibinfo {author} {\bibfnamefont {A.~M.}\ \bibnamefont
  {Yao}},\ }\bibfield  {title} {\bibinfo {title} {Chirality and the angular
  momentum of light},\ }\href {https://doi.org/10.1098/rsta.2015.0433}
  {\bibfield  {journal} {\bibinfo  {journal} {Philosophical Transactions of the
  Royal Society A: Mathematical, Physical and Engineering Sciences}\ }\textbf
  {\bibinfo {volume} {375}},\ \bibinfo {pages} {20150433} (\bibinfo {year}
  {2017})}\BibitemShut {NoStop}%
\bibitem [{\citenamefont {V{\'a}zquez-Lozano}\ and\ \citenamefont
  {Mart{\'\i}nez}(2018)}]{VazquezLozano2018}%
  \BibitemOpen
  \bibfield  {author} {\bibinfo {author} {\bibfnamefont {J.~E.}\ \bibnamefont
  {V{\'a}zquez-Lozano}}\ and\ \bibinfo {author} {\bibfnamefont
  {A.}~\bibnamefont {Mart{\'\i}nez}},\ }\bibfield  {title} {\bibinfo {title}
  {Optical chirality in dispersive and lossy media},\ }\href
  {https://doi.org/10.1103/PhysRevLett.121.043901} {\bibfield  {journal}
  {\bibinfo  {journal} {Physical Review Letters}\ }\textbf {\bibinfo {volume}
  {121}},\ \bibinfo {pages} {043901} (\bibinfo {year} {2018})}\BibitemShut
  {NoStop}%
\bibitem [{\citenamefont {Nienhuis}(2016)}]{Nienhuis2016}%
  \BibitemOpen
  \bibfield  {author} {\bibinfo {author} {\bibfnamefont {G.}~\bibnamefont
  {Nienhuis}},\ }\bibfield  {title} {\bibinfo {title} {Conservation laws and
  symmetry transformations of the electromagnetic field with sources},\ }\href
  {https://doi.org/10.1103/PhysRevA.93.023840} {\bibfield  {journal} {\bibinfo
  {journal} {Physical Review A}\ }\textbf {\bibinfo {volume} {93}},\ \bibinfo
  {pages} {023840} (\bibinfo {year} {2016})}\BibitemShut {NoStop}%
\bibitem [{\citenamefont {Fernandez-Corbaton}\ and\ \citenamefont
  {Rockstuhl}(2017)}]{FernandezCorbaton2017}%
  \BibitemOpen
  \bibfield  {author} {\bibinfo {author} {\bibfnamefont {I.}~\bibnamefont
  {Fernandez-Corbaton}}\ and\ \bibinfo {author} {\bibfnamefont
  {C.}~\bibnamefont {Rockstuhl}},\ }\bibfield  {title} {\bibinfo {title}
  {Unified theory to describe and engineer conservation laws in light-matter
  interactions},\ }\href {https://doi.org/10.1103/PhysRevA.95.053829}
  {\bibfield  {journal} {\bibinfo  {journal} {Physical Review A}\ }\textbf
  {\bibinfo {volume} {95}},\ \bibinfo {pages} {053829} (\bibinfo {year}
  {2017})}\BibitemShut {NoStop}%
\bibitem [{\citenamefont {Lobet}\ \emph {et~al.}(2022)\citenamefont {Lobet},
  \citenamefont {Liberal}, \citenamefont {Vertchenko}, \citenamefont
  {Lavrinenko}, \citenamefont {Engheta},\ and\ \citenamefont
  {Mazur}}]{lobet2022momentum}%
  \BibitemOpen
  \bibfield  {author} {\bibinfo {author} {\bibfnamefont {M.}~\bibnamefont
  {Lobet}}, \bibinfo {author} {\bibfnamefont {I.}~\bibnamefont {Liberal}},
  \bibinfo {author} {\bibfnamefont {L.}~\bibnamefont {Vertchenko}}, \bibinfo
  {author} {\bibfnamefont {A.~V.}\ \bibnamefont {Lavrinenko}}, \bibinfo
  {author} {\bibfnamefont {N.}~\bibnamefont {Engheta}},\ and\ \bibinfo {author}
  {\bibfnamefont {E.}~\bibnamefont {Mazur}},\ }\bibfield  {title} {\bibinfo
  {title} {Momentum considerations inside near-zero index materials},\
  }\href@noop {} {\bibfield  {journal} {\bibinfo  {journal} {Light: Science \&
  Applications}\ }\textbf {\bibinfo {volume} {11}},\ \bibinfo {pages} {110}
  (\bibinfo {year} {2022})}\BibitemShut {NoStop}%
\bibitem [{\citenamefont {Barnett}\ and\ \citenamefont
  {Loudon}(2010)}]{Barnett2010A}%
  \BibitemOpen
  \bibfield  {author} {\bibinfo {author} {\bibfnamefont {S.~M.}\ \bibnamefont
  {Barnett}}\ and\ \bibinfo {author} {\bibfnamefont {R.}~\bibnamefont
  {Loudon}},\ }\bibfield  {title} {\bibinfo {title} {The enigma of optical
  momentum in a medium},\ }\href {https://doi.org/10.1098/rsta.2009.0207}
  {\bibfield  {journal} {\bibinfo  {journal} {Philosophical Transactions of the
  Royal Society A: Mathematical, Physical and Engineering Sciences}\ }\textbf
  {\bibinfo {volume} {368}},\ \bibinfo {pages} {927} (\bibinfo {year}
  {2010})}\BibitemShut {NoStop}%
\bibitem [{\citenamefont {Barnett}(2010)}]{Barnett2010B}%
  \BibitemOpen
  \bibfield  {author} {\bibinfo {author} {\bibfnamefont {S.~M.}\ \bibnamefont
  {Barnett}},\ }\bibfield  {title} {\bibinfo {title} {Resolution of the
  abraham-minkowski dilemma},\ }\href
  {https://doi.org/10.1103/PhysRevLett.104.070401} {\bibfield  {journal}
  {\bibinfo  {journal} {Physical review letters}\ }\textbf {\bibinfo {volume}
  {104}},\ \bibinfo {pages} {070401} (\bibinfo {year} {2010})}\BibitemShut
  {NoStop}%
\bibitem [{\citenamefont {Silveirinha}(2017)}]{Silveirinha2017}%
  \BibitemOpen
  \bibfield  {author} {\bibinfo {author} {\bibfnamefont {M.~G.}\ \bibnamefont
  {Silveirinha}},\ }\bibfield  {title} {\bibinfo {title} {Reexamination of the
  abraham-minkowski dilemma},\ }\href
  {https://doi.org/10.1103/PhysRevA.96.033831} {\bibfield  {journal} {\bibinfo
  {journal} {Physical Review A}\ }\textbf {\bibinfo {volume} {96}},\ \bibinfo
  {pages} {033831} (\bibinfo {year} {2017})}\BibitemShut {NoStop}%
\bibitem [{\citenamefont {Philbin}(2013)}]{Philbin2013}%
  \BibitemOpen
  \bibfield  {author} {\bibinfo {author} {\bibfnamefont {T.~G.}\ \bibnamefont
  {Philbin}},\ }\bibfield  {title} {\bibinfo {title} {Lipkin's conservation
  law, noether's theorem, and the relation to optical helicity},\ }\href
  {https://doi.org/10.1103/PhysRevA.96.033831} {\bibfield  {journal} {\bibinfo
  {journal} {Physical Review A}\ }\textbf {\bibinfo {volume} {87}},\ \bibinfo
  {pages} {043843} (\bibinfo {year} {2013})}\BibitemShut {NoStop}%
\bibitem [{\citenamefont {Crimin}\ \emph {et~al.}(2019)\citenamefont {Crimin},
  \citenamefont {Mackinnon}, \citenamefont {G{\"o}tte},\ and\ \citenamefont
  {Barnett}}]{Crimin2019}%
  \BibitemOpen
  \bibfield  {author} {\bibinfo {author} {\bibfnamefont {F.}~\bibnamefont
  {Crimin}}, \bibinfo {author} {\bibfnamefont {N.}~\bibnamefont {Mackinnon}},
  \bibinfo {author} {\bibfnamefont {J.~B.}\ \bibnamefont {G{\"o}tte}},\ and\
  \bibinfo {author} {\bibfnamefont {S.~M.}\ \bibnamefont {Barnett}},\
  }\bibfield  {title} {\bibinfo {title} {Optical helicity and chirality:
  conservation and sources},\ }\href {https://doi.org/10.3390/app9050828}
  {\bibfield  {journal} {\bibinfo  {journal} {Applied Sciences}\ }\textbf
  {\bibinfo {volume} {9}},\ \bibinfo {pages} {828} (\bibinfo {year}
  {2019})}\BibitemShut {NoStop}%
\bibitem [{\citenamefont {Mackinnon}(2019)}]{Mackinnon2019}%
  \BibitemOpen
  \bibfield  {author} {\bibinfo {author} {\bibfnamefont {N.}~\bibnamefont
  {Mackinnon}},\ }\bibfield  {title} {\bibinfo {title} {On the differences
  between helicity and chirality},\ }\href
  {https://doi.org/10.1088/2040-8986/ab4db9} {\bibfield  {journal} {\bibinfo
  {journal} {Journal of Optics}\ }\textbf {\bibinfo {volume} {21}},\ \bibinfo
  {pages} {125402} (\bibinfo {year} {2019})}\BibitemShut {NoStop}%
\bibitem [{\citenamefont {Poulikakos}\ \emph {et~al.}(2019)\citenamefont
  {Poulikakos}, \citenamefont {Dionne},\ and\ \citenamefont
  {Garc{\'\i}a-Etxarri}}]{Poulikakos2019}%
  \BibitemOpen
  \bibfield  {author} {\bibinfo {author} {\bibfnamefont {L.~V.}\ \bibnamefont
  {Poulikakos}}, \bibinfo {author} {\bibfnamefont {J.~A.}\ \bibnamefont
  {Dionne}},\ and\ \bibinfo {author} {\bibfnamefont {A.}~\bibnamefont
  {Garc{\'\i}a-Etxarri}},\ }\bibfield  {title} {\bibinfo {title} {Optical
  helicity and optical chirality in free space and in the presence of matter},\
  }\href {https://doi.org/10.3390/sym11091113} {\bibfield  {journal} {\bibinfo
  {journal} {Symmetry}\ }\textbf {\bibinfo {volume} {11}},\ \bibinfo {pages}
  {1113} (\bibinfo {year} {2019})}\BibitemShut {NoStop}%
\bibitem [{\citenamefont {Hayran}\ \emph {et~al.}(2022)\citenamefont {Hayran},
  \citenamefont {Khurgin},\ and\ \citenamefont {Monticone}}]{hayran2022}%
  \BibitemOpen
  \bibfield  {author} {\bibinfo {author} {\bibfnamefont {Z.}~\bibnamefont
  {Hayran}}, \bibinfo {author} {\bibfnamefont {J.~B.}\ \bibnamefont
  {Khurgin}},\ and\ \bibinfo {author} {\bibfnamefont {F.}~\bibnamefont
  {Monticone}},\ }\bibfield  {title} {\bibinfo {title} {$\hbar$$\omega$ versus
  $\hbar$k: dispersion and energy constraints on time-varying photonic
  materials and time crystals},\ }\href {https://doi.org/10.1364/OME.471672}
  {\bibfield  {journal} {\bibinfo  {journal} {Optical Materials Express}\
  }\textbf {\bibinfo {volume} {12}},\ \bibinfo {pages} {3904} (\bibinfo {year}
  {2022})}\BibitemShut {NoStop}%
\bibitem [{\citenamefont {Sol\'{i}s}\ \emph {et~al.}(2021)\citenamefont
  {Sol\'{i}s}, \citenamefont {Kastner},\ and\ \citenamefont {Engheta}}]{Solis}%
  \BibitemOpen
  \bibfield  {author} {\bibinfo {author} {\bibfnamefont {D.~M.}\ \bibnamefont
  {Sol\'{i}s}}, \bibinfo {author} {\bibfnamefont {R.}~\bibnamefont {Kastner}},\
  and\ \bibinfo {author} {\bibfnamefont {N.}~\bibnamefont {Engheta}},\
  }\bibfield  {title} {\bibinfo {title} {Time-varying materials in the presence
  of dispersion: plane-wave propagation in a lorentzian medium with temporal
  discontinuity},\ }\href {https://doi.org/10.1364/PRJ.427368} {\bibfield
  {journal} {\bibinfo  {journal} {Photon. Res.}\ }\textbf {\bibinfo {volume}
  {9}},\ \bibinfo {pages} {1842} (\bibinfo {year} {2021})}\BibitemShut
  {NoStop}%
\bibitem [{\citenamefont {Gratus}\ \emph {et~al.}(2021)\citenamefont {Gratus},
  \citenamefont {Seviour}, \citenamefont {Kinsler},\ and\ \citenamefont
  {Jaroszynski}}]{gratus2021}%
  \BibitemOpen
  \bibfield  {author} {\bibinfo {author} {\bibfnamefont {J.}~\bibnamefont
  {Gratus}}, \bibinfo {author} {\bibfnamefont {R.}~\bibnamefont {Seviour}},
  \bibinfo {author} {\bibfnamefont {P.}~\bibnamefont {Kinsler}},\ and\ \bibinfo
  {author} {\bibfnamefont {D.~A.}\ \bibnamefont {Jaroszynski}},\ }\bibfield
  {title} {\bibinfo {title} {Temporal boundaries in electromagnetic
  materials},\ }\href {https://doi.org/10.1088/1367-2630/ac1896} {\bibfield
  {journal} {\bibinfo  {journal} {New Journal of Physics}\ }\textbf {\bibinfo
  {volume} {23}},\ \bibinfo {pages} {083032} (\bibinfo {year}
  {2021})}\BibitemShut {NoStop}%
\bibitem [{\citenamefont {Mai}\ \emph {et~al.}(2023)\citenamefont {Mai},
  \citenamefont {Xu},\ and\ \citenamefont {Werner}}]{mai2023}%
  \BibitemOpen
  \bibfield  {author} {\bibinfo {author} {\bibfnamefont {W.}~\bibnamefont
  {Mai}}, \bibinfo {author} {\bibfnamefont {J.}~\bibnamefont {Xu}},\ and\
  \bibinfo {author} {\bibfnamefont {D.~H.}\ \bibnamefont {Werner}},\ }\bibfield
   {title} {\bibinfo {title} {Fundamental asymmetries between spatial and
  temporal boundaries in electromagnetics},\ }\href
  {https://doi.org/10.3390/sym15040858} {\bibfield  {journal} {\bibinfo
  {journal} {Symmetry}\ }\textbf {\bibinfo {volume} {15}},\ \bibinfo {pages}
  {858} (\bibinfo {year} {2023})}\BibitemShut {NoStop}%
\bibitem [{\citenamefont {Zurita-S{\'a}nchez}\ \emph
  {et~al.}(2009)\citenamefont {Zurita-S{\'a}nchez}, \citenamefont {Halevi},\
  and\ \citenamefont {Cervantes-Gonzalez}}]{zurita2009}%
  \BibitemOpen
  \bibfield  {author} {\bibinfo {author} {\bibfnamefont {J.~R.}\ \bibnamefont
  {Zurita-S{\'a}nchez}}, \bibinfo {author} {\bibfnamefont {P.}~\bibnamefont
  {Halevi}},\ and\ \bibinfo {author} {\bibfnamefont {J.~C.}\ \bibnamefont
  {Cervantes-Gonzalez}},\ }\bibfield  {title} {\bibinfo {title} {Reflection and
  transmission of a wave incident on a slab with a time-periodic dielectric
  function},\ }\href {https://doi.org/10.1103/PhysRevA.79.053821} {\bibfield
  {journal} {\bibinfo  {journal} {Physical Review A}\ }\textbf {\bibinfo
  {volume} {79}},\ \bibinfo {pages} {053821} (\bibinfo {year}
  {2009})}\BibitemShut {NoStop}%
\bibitem [{\citenamefont {Gaxiola-Luna}\ and\ \citenamefont
  {Halevi}(2021)}]{gaxiola2021}%
  \BibitemOpen
  \bibfield  {author} {\bibinfo {author} {\bibfnamefont {J.~G.}\ \bibnamefont
  {Gaxiola-Luna}}\ and\ \bibinfo {author} {\bibfnamefont {P.}~\bibnamefont
  {Halevi}},\ }\bibfield  {title} {\bibinfo {title} {Temporal photonic (time)
  crystal with a square profile of both permittivity ɛ (t) and permeability
  $\mu$ (t)},\ }\href {https://doi.org/10.1103/PhysRevB.103.144306} {\bibfield
  {journal} {\bibinfo  {journal} {Physical Review B}\ }\textbf {\bibinfo
  {volume} {103}},\ \bibinfo {pages} {144306} (\bibinfo {year}
  {2021})}\BibitemShut {NoStop}%
\bibitem [{\citenamefont {Ramaccia}\ \emph {et~al.}(2020)\citenamefont
  {Ramaccia}, \citenamefont {Sounas}, \citenamefont {Marini}, \citenamefont
  {Toscano},\ and\ \citenamefont {Bilotti}}]{ramaccia2020}%
  \BibitemOpen
  \bibfield  {author} {\bibinfo {author} {\bibfnamefont {D.}~\bibnamefont
  {Ramaccia}}, \bibinfo {author} {\bibfnamefont {D.~L.}\ \bibnamefont
  {Sounas}}, \bibinfo {author} {\bibfnamefont {A.~V.}\ \bibnamefont {Marini}},
  \bibinfo {author} {\bibfnamefont {A.}~\bibnamefont {Toscano}},\ and\ \bibinfo
  {author} {\bibfnamefont {F.}~\bibnamefont {Bilotti}},\ }\bibfield  {title}
  {\bibinfo {title} {Electromagnetic isolation induced by time-varying
  metasurfaces: Nonreciprocal bragg grating},\ }\href
  {https://doi.org/10.1109/LAWP.2020.2996275} {\bibfield  {journal} {\bibinfo
  {journal} {IEEE Antennas and Wireless Propagation Letters}\ }\textbf
  {\bibinfo {volume} {19}},\ \bibinfo {pages} {1886} (\bibinfo {year}
  {2020})}\BibitemShut {NoStop}%
\bibitem [{\citenamefont {Ramaccia}\ \emph {et~al.}(2021)\citenamefont
  {Ramaccia}, \citenamefont {Al{\`u}}, \citenamefont {Toscano},\ and\
  \citenamefont {Bilotti}}]{ramaccia2021}%
  \BibitemOpen
  \bibfield  {author} {\bibinfo {author} {\bibfnamefont {D.}~\bibnamefont
  {Ramaccia}}, \bibinfo {author} {\bibfnamefont {A.}~\bibnamefont {Al{\`u}}},
  \bibinfo {author} {\bibfnamefont {A.}~\bibnamefont {Toscano}},\ and\ \bibinfo
  {author} {\bibfnamefont {F.}~\bibnamefont {Bilotti}},\ }\bibfield  {title}
  {\bibinfo {title} {Temporal multilayer structures for designing higher-order
  transfer functions using time-varying metamaterials},\ }\bibfield  {journal}
  {\bibinfo  {journal} {Applied Physics Letters}\ }\textbf {\bibinfo {volume}
  {118}},\ \href {https://doi.org/10.1063/5.0042567} {10.1063/5.0042567}
  (\bibinfo {year} {2021})\BibitemShut {NoStop}%
\bibitem [{\citenamefont {Kong}(1975)}]{kong1975}%
  \BibitemOpen
  \bibfield  {author} {\bibinfo {author} {\bibfnamefont {J.~A.}\ \bibnamefont
  {Kong}},\ }\bibfield  {title} {\bibinfo {title} {Theory of electromagnetic
  waves},\ }\href@noop {} {\bibfield  {journal} {\bibinfo  {journal} {New
  York}\ } (\bibinfo {year} {1975})}\BibitemShut {NoStop}%
\bibitem [{\citenamefont {Bliokh}\ \emph {et~al.}(2013)\citenamefont {Bliokh},
  \citenamefont {Bekshaev},\ and\ \citenamefont {Nori}}]{bliokh2013dual}%
  \BibitemOpen
  \bibfield  {author} {\bibinfo {author} {\bibfnamefont {K.~Y.}\ \bibnamefont
  {Bliokh}}, \bibinfo {author} {\bibfnamefont {A.~Y.}\ \bibnamefont
  {Bekshaev}},\ and\ \bibinfo {author} {\bibfnamefont {F.}~\bibnamefont
  {Nori}},\ }\bibfield  {title} {\bibinfo {title} {Dual electromagnetism:
  helicity, spin, momentum and angular momentum},\ }\href
  {https://doi.org/10.1088/1367-2630/15/3/033026} {\bibfield  {journal}
  {\bibinfo  {journal} {New Journal of Physics}\ }\textbf {\bibinfo {volume}
  {15}},\ \bibinfo {pages} {033026} (\bibinfo {year} {2013})}\BibitemShut
  {NoStop}%
\bibitem [{\citenamefont {Berry}(2009)}]{berry2009optical}%
  \BibitemOpen
  \bibfield  {author} {\bibinfo {author} {\bibfnamefont {M.~V.}\ \bibnamefont
  {Berry}},\ }\bibfield  {title} {\bibinfo {title} {Optical currents},\ }\href
  {https://doi.org/10.1088/1464-4258/11/9/094001} {\bibfield  {journal}
  {\bibinfo  {journal} {Journal of Optics A: Pure and Applied Optics}\ }\textbf
  {\bibinfo {volume} {11}},\ \bibinfo {pages} {094001} (\bibinfo {year}
  {2009})}\BibitemShut {NoStop}%
\bibitem [{\citenamefont {Avetisyan}\ \emph {et~al.}(2021)\citenamefont
  {Avetisyan}, \citenamefont {Evnin},\ and\ \citenamefont
  {Mkrtchyan}}]{avetisyan2021democratic}%
  \BibitemOpen
  \bibfield  {author} {\bibinfo {author} {\bibfnamefont {Z.}~\bibnamefont
  {Avetisyan}}, \bibinfo {author} {\bibfnamefont {O.}~\bibnamefont {Evnin}},\
  and\ \bibinfo {author} {\bibfnamefont {K.}~\bibnamefont {Mkrtchyan}},\
  }\bibfield  {title} {\bibinfo {title} {Democratic lagrangians for nonlinear
  electrodynamics},\ }\href {https://doi.org/10.1103/PhysRevLett.127.271601}
  {\bibfield  {journal} {\bibinfo  {journal} {Physical Review Letters}\
  }\textbf {\bibinfo {volume} {127}},\ \bibinfo {pages} {271601} (\bibinfo
  {year} {2021})}\BibitemShut {NoStop}%
\bibitem [{\citenamefont {Bliokh}\ \emph
  {et~al.}(2015{\natexlab{a}})\citenamefont {Bliokh}, \citenamefont
  {Rodr{\'\i}guez-Fortu{\~n}o}, \citenamefont {Nori},\ and\ \citenamefont
  {Zayats}}]{bliokh2015}%
  \BibitemOpen
  \bibfield  {author} {\bibinfo {author} {\bibfnamefont {K.~Y.}\ \bibnamefont
  {Bliokh}}, \bibinfo {author} {\bibfnamefont {F.~J.}\ \bibnamefont
  {Rodr{\'\i}guez-Fortu{\~n}o}}, \bibinfo {author} {\bibfnamefont
  {F.}~\bibnamefont {Nori}},\ and\ \bibinfo {author} {\bibfnamefont {A.~V.}\
  \bibnamefont {Zayats}},\ }\bibfield  {title} {\bibinfo {title} {Spin--orbit
  interactions of light},\ }\href {https://doi.org/10.1038/nphoton.2015.201}
  {\bibfield  {journal} {\bibinfo  {journal} {Nature Photonics}\ }\textbf
  {\bibinfo {volume} {9}},\ \bibinfo {pages} {796} (\bibinfo {year}
  {2015}{\natexlab{a}})}\BibitemShut {NoStop}%
\bibitem [{\citenamefont {Haefner}\ \emph {et~al.}(2009)\citenamefont
  {Haefner}, \citenamefont {Sukhov},\ and\ \citenamefont
  {Dogariu}}]{haefner2009}%
  \BibitemOpen
  \bibfield  {author} {\bibinfo {author} {\bibfnamefont {D.}~\bibnamefont
  {Haefner}}, \bibinfo {author} {\bibfnamefont {S.}~\bibnamefont {Sukhov}},\
  and\ \bibinfo {author} {\bibfnamefont {A.}~\bibnamefont {Dogariu}},\
  }\bibfield  {title} {\bibinfo {title} {Spin hall effect of light in spherical
  geometry},\ }\href {https://doi.org/10.1103/PhysRevLett.102.123903}
  {\bibfield  {journal} {\bibinfo  {journal} {Physical review letters}\
  }\textbf {\bibinfo {volume} {102}},\ \bibinfo {pages} {123903} (\bibinfo
  {year} {2009})}\BibitemShut {NoStop}%
\bibitem [{\citenamefont {Bliokh}\ \emph
  {et~al.}(2015{\natexlab{b}})\citenamefont {Bliokh}, \citenamefont
  {Smirnova},\ and\ \citenamefont {Nori}}]{bliokh2015q}%
  \BibitemOpen
  \bibfield  {author} {\bibinfo {author} {\bibfnamefont {K.~Y.}\ \bibnamefont
  {Bliokh}}, \bibinfo {author} {\bibfnamefont {D.}~\bibnamefont {Smirnova}},\
  and\ \bibinfo {author} {\bibfnamefont {F.}~\bibnamefont {Nori}},\ }\bibfield
  {title} {\bibinfo {title} {Quantum spin hall effect of light},\ }\href
  {https://doi.org/10.1126/science.aaa9519} {\bibfield  {journal} {\bibinfo
  {journal} {Science}\ }\textbf {\bibinfo {volume} {348}},\ \bibinfo {pages}
  {1448} (\bibinfo {year} {2015}{\natexlab{b}})}\BibitemShut {NoStop}%
\bibitem [{\citenamefont {Van~Mechelen}\ and\ \citenamefont
  {Jacob}(2016)}]{van2016}%
  \BibitemOpen
  \bibfield  {author} {\bibinfo {author} {\bibfnamefont {T.}~\bibnamefont
  {Van~Mechelen}}\ and\ \bibinfo {author} {\bibfnamefont {Z.}~\bibnamefont
  {Jacob}},\ }\bibfield  {title} {\bibinfo {title} {Universal spin-momentum
  locking of evanescent waves},\ }\href
  {https://doi.org/10.1364/OPTICA.3.000118} {\bibfield  {journal} {\bibinfo
  {journal} {Optica}\ }\textbf {\bibinfo {volume} {3}},\ \bibinfo {pages} {118}
  (\bibinfo {year} {2016})}\BibitemShut {NoStop}%
\bibitem [{\citenamefont {Trueba}\ and\ \citenamefont {Ranada}(1996)}]{trueba}%
  \BibitemOpen
  \bibfield  {author} {\bibinfo {author} {\bibfnamefont {J.~L.}\ \bibnamefont
  {Trueba}}\ and\ \bibinfo {author} {\bibfnamefont {A.~F.}\ \bibnamefont
  {Ranada}},\ }\bibfield  {title} {\bibinfo {title} {The electromagnetic
  helicity},\ }\href {https://doi.org/10.1088/0143-0807/17/3/008} {\bibfield
  {journal} {\bibinfo  {journal} {European Journal of Physics}\ }\textbf
  {\bibinfo {volume} {17}},\ \bibinfo {pages} {141} (\bibinfo {year}
  {1996})}\BibitemShut {NoStop}%
\bibitem [{\citenamefont {Negoro}\ \emph {et~al.}(2023)\citenamefont {Negoro},
  \citenamefont {Sugimoto},\ and\ \citenamefont {Fujii}}]{negoro2023}%
  \BibitemOpen
  \bibfield  {author} {\bibinfo {author} {\bibfnamefont {H.}~\bibnamefont
  {Negoro}}, \bibinfo {author} {\bibfnamefont {H.}~\bibnamefont {Sugimoto}},\
  and\ \bibinfo {author} {\bibfnamefont {M.}~\bibnamefont {Fujii}},\ }\bibfield
   {title} {\bibinfo {title} {Helicity-preserving optical metafluids},\ }\href
  {https://doi.org/10.1021/acs.nanolett.3c01026} {\bibfield  {journal}
  {\bibinfo  {journal} {Nano Letters}\ }\textbf {\bibinfo {volume} {23}},\
  \bibinfo {pages} {5101} (\bibinfo {year} {2023})}\BibitemShut {NoStop}%
\bibitem [{\citenamefont {Matyushkin}\ \emph {et~al.}(2020)\citenamefont
  {Matyushkin}, \citenamefont {Danilov}, \citenamefont {Moskotin},
  \citenamefont {Belosevich}, \citenamefont {Kaurova}, \citenamefont {Rybin},
  \citenamefont {Obraztsova}, \citenamefont {Fedorov}, \citenamefont
  {Gorbenko}, \citenamefont {Kachorovskii} \emph {et~al.}}]{matyushkin2020}%
  \BibitemOpen
  \bibfield  {author} {\bibinfo {author} {\bibfnamefont {Y.}~\bibnamefont
  {Matyushkin}}, \bibinfo {author} {\bibfnamefont {S.}~\bibnamefont {Danilov}},
  \bibinfo {author} {\bibfnamefont {M.}~\bibnamefont {Moskotin}}, \bibinfo
  {author} {\bibfnamefont {V.}~\bibnamefont {Belosevich}}, \bibinfo {author}
  {\bibfnamefont {N.}~\bibnamefont {Kaurova}}, \bibinfo {author} {\bibfnamefont
  {M.}~\bibnamefont {Rybin}}, \bibinfo {author} {\bibfnamefont {E.~D.}\
  \bibnamefont {Obraztsova}}, \bibinfo {author} {\bibfnamefont
  {G.}~\bibnamefont {Fedorov}}, \bibinfo {author} {\bibfnamefont
  {I.}~\bibnamefont {Gorbenko}}, \bibinfo {author} {\bibfnamefont
  {V.}~\bibnamefont {Kachorovskii}}, \emph {et~al.},\ }\bibfield  {title}
  {\bibinfo {title} {Helicity-sensitive plasmonic terahertz interferometer},\
  }\href {https://doi.org/10.1021/acs.nanolett.3c01026} {\bibfield  {journal}
  {\bibinfo  {journal} {Nano Letters}\ }\textbf {\bibinfo {volume} {20}},\
  \bibinfo {pages} {7296} (\bibinfo {year} {2020})}\BibitemShut {NoStop}%
\bibitem [{\citenamefont {Olmos-Trigo}\ and\ \citenamefont
  {Zambrana-Puyalto}(2022)}]{olmos2022helicity}%
  \BibitemOpen
  \bibfield  {author} {\bibinfo {author} {\bibfnamefont {J.}~\bibnamefont
  {Olmos-Trigo}}\ and\ \bibinfo {author} {\bibfnamefont {X.}~\bibnamefont
  {Zambrana-Puyalto}},\ }\bibfield  {title} {\bibinfo {title} {Helicity
  conservation for mie optical cavities},\ }\href
  {https://doi.org/10.1103/PhysRevApplied.18.044007} {\bibfield  {journal}
  {\bibinfo  {journal} {Physical Review Applied}\ }\textbf {\bibinfo {volume}
  {18}},\ \bibinfo {pages} {044007} (\bibinfo {year} {2022})}\BibitemShut
  {NoStop}%
\bibitem [{\citenamefont {Balanis}(2012)}]{balanis}%
  \BibitemOpen
  \bibfield  {author} {\bibinfo {author} {\bibfnamefont {C.~A.}\ \bibnamefont
  {Balanis}},\ }\href@noop {} {\emph {\bibinfo {title} {Advanced engineering
  electromagnetics}}}\ (\bibinfo  {publisher} {John Wiley \& Sons},\ \bibinfo
  {year} {2012})\BibitemShut {NoStop}%
\bibitem [{\citenamefont {Bliokh}\ and\ \citenamefont
  {Nori}(2011)}]{bliokh2011characterizing}%
  \BibitemOpen
  \bibfield  {author} {\bibinfo {author} {\bibfnamefont {K.~Y.}\ \bibnamefont
  {Bliokh}}\ and\ \bibinfo {author} {\bibfnamefont {F.}~\bibnamefont {Nori}},\
  }\bibfield  {title} {\bibinfo {title} {Characterizing optical chirality},\
  }\href {https://doi.org/10.1103/PhysRevA.83.021803} {\bibfield  {journal}
  {\bibinfo  {journal} {Physical Review A}\ }\textbf {\bibinfo {volume} {83}},\
  \bibinfo {pages} {021803} (\bibinfo {year} {2011})}\BibitemShut {NoStop}%
\bibitem [{\citenamefont {Yoo}\ and\ \citenamefont {Park}(2019)}]{yoo}%
  \BibitemOpen
  \bibfield  {author} {\bibinfo {author} {\bibfnamefont {S.}~\bibnamefont
  {Yoo}}\ and\ \bibinfo {author} {\bibfnamefont {Q.-H.}\ \bibnamefont {Park}},\
  }\bibfield  {title} {\bibinfo {title} {Metamaterials and chiral sensing: a
  review of fundamentals and applications},\ }\href
  {https://doi.org/10.1515/nanoph-2018-0167} {\bibfield  {journal} {\bibinfo
  {journal} {Nanophotonics}\ }\textbf {\bibinfo {volume} {8}},\ \bibinfo
  {pages} {249} (\bibinfo {year} {2019})}\BibitemShut {NoStop}%
\bibitem [{\citenamefont {Formen}\ \emph {et~al.}(2024)\citenamefont {Formen},
  \citenamefont {Howard}, \citenamefont {Anslyn},\ and\ \citenamefont
  {Wolf}}]{formen2024}%
  \BibitemOpen
  \bibfield  {author} {\bibinfo {author} {\bibfnamefont {J.~S.}\ \bibnamefont
  {Formen}}, \bibinfo {author} {\bibfnamefont {J.~R.}\ \bibnamefont {Howard}},
  \bibinfo {author} {\bibfnamefont {E.~V.}\ \bibnamefont {Anslyn}},\ and\
  \bibinfo {author} {\bibfnamefont {C.}~\bibnamefont {Wolf}},\ }\bibfield
  {title} {\bibinfo {title} {Circular dichroism sensing: Strategies and
  applications},\ }\href {https://doi.org/10.1002/anie.202400767} {\bibfield
  {journal} {\bibinfo  {journal} {Angewandte Chemie International Edition}\ ,\
  \bibinfo {pages} {e202400767}} (\bibinfo {year} {2024})}\BibitemShut
  {NoStop}%
\bibitem [{\citenamefont {Olmos-Trigo}\ \emph {et~al.}(2024)\citenamefont
  {Olmos-Trigo}, \citenamefont {Lasa-Alonso}, \citenamefont
  {G{\'o}mez-Viloria}, \citenamefont {Molina-Terriza},\ and\ \citenamefont
  {Garc{\'\i}a-Etxarri}}]{olmos2024capturing}%
  \BibitemOpen
  \bibfield  {author} {\bibinfo {author} {\bibfnamefont {J.}~\bibnamefont
  {Olmos-Trigo}}, \bibinfo {author} {\bibfnamefont {J.}~\bibnamefont
  {Lasa-Alonso}}, \bibinfo {author} {\bibfnamefont {I.}~\bibnamefont
  {G{\'o}mez-Viloria}}, \bibinfo {author} {\bibfnamefont {G.}~\bibnamefont
  {Molina-Terriza}},\ and\ \bibinfo {author} {\bibfnamefont {A.}~\bibnamefont
  {Garc{\'\i}a-Etxarri}},\ }\bibfield  {title} {\bibinfo {title} {Capturing
  near-field circular dichroism enhancements from far-field measurements},\
  }\href {https://doi.org/10.1103/PhysRevResearch.6.013151} {\bibfield
  {journal} {\bibinfo  {journal} {Physical Review Research}\ }\textbf {\bibinfo
  {volume} {6}},\ \bibinfo {pages} {013151} (\bibinfo {year}
  {2024})}\BibitemShut {NoStop}%
\bibitem [{\citenamefont {Huang}\ \emph {et~al.}(2024)\citenamefont {Huang},
  \citenamefont {Chen}, \citenamefont {Ma}, \citenamefont {Huang},
  \citenamefont {Zhi}, \citenamefont {Li}, \citenamefont {Zeng}, \citenamefont
  {Pi}, \citenamefont {Xu}, \citenamefont {Xu} \emph {et~al.}}]{huang2024}%
  \BibitemOpen
  \bibfield  {author} {\bibinfo {author} {\bibfnamefont {X.}~\bibnamefont
  {Huang}}, \bibinfo {author} {\bibfnamefont {Q.}~\bibnamefont {Chen}},
  \bibinfo {author} {\bibfnamefont {Y.}~\bibnamefont {Ma}}, \bibinfo {author}
  {\bibfnamefont {C.}~\bibnamefont {Huang}}, \bibinfo {author} {\bibfnamefont
  {W.}~\bibnamefont {Zhi}}, \bibinfo {author} {\bibfnamefont {J.}~\bibnamefont
  {Li}}, \bibinfo {author} {\bibfnamefont {R.}~\bibnamefont {Zeng}}, \bibinfo
  {author} {\bibfnamefont {J.}~\bibnamefont {Pi}}, \bibinfo {author}
  {\bibfnamefont {J.-f.}\ \bibnamefont {Xu}}, \bibinfo {author} {\bibfnamefont
  {J.}~\bibnamefont {Xu}}, \emph {et~al.},\ }\bibfield  {title} {\bibinfo
  {title} {Chiral au nanostars for sers sensing of enantiomers discrimination,
  multibacteria recognition and photothermal antibacterial application},\
  }\href {https://doi.org/10.1016/j.cej.2023.147528} {\bibfield  {journal}
  {\bibinfo  {journal} {Chemical Engineering Journal}\ }\textbf {\bibinfo
  {volume} {479}},\ \bibinfo {pages} {147528} (\bibinfo {year}
  {2024})}\BibitemShut {NoStop}%
\bibitem [{\citenamefont {Hendry}\ \emph {et~al.}(2010)\citenamefont {Hendry},
  \citenamefont {Carpy}, \citenamefont {Johnston}, \citenamefont {Popland},
  \citenamefont {Mikhaylovskiy}, \citenamefont {Lapthorn}, \citenamefont
  {Kelly}, \citenamefont {Barron}, \citenamefont {Gadegaard},\ and\
  \citenamefont {Kadodwala}}]{Hendry2010}%
  \BibitemOpen
  \bibfield  {author} {\bibinfo {author} {\bibfnamefont {E.}~\bibnamefont
  {Hendry}}, \bibinfo {author} {\bibfnamefont {T.}~\bibnamefont {Carpy}},
  \bibinfo {author} {\bibfnamefont {J.}~\bibnamefont {Johnston}}, \bibinfo
  {author} {\bibfnamefont {M.}~\bibnamefont {Popland}}, \bibinfo {author}
  {\bibfnamefont {R.~V.}\ \bibnamefont {Mikhaylovskiy}}, \bibinfo {author}
  {\bibfnamefont {A.~J.}\ \bibnamefont {Lapthorn}}, \bibinfo {author}
  {\bibfnamefont {S.~M.}\ \bibnamefont {Kelly}}, \bibinfo {author}
  {\bibfnamefont {L.~D.}\ \bibnamefont {Barron}}, \bibinfo {author}
  {\bibfnamefont {N.}~\bibnamefont {Gadegaard}},\ and\ \bibinfo {author}
  {\bibfnamefont {M.}~\bibnamefont {Kadodwala}},\ }\bibfield  {title} {\bibinfo
  {title} {Ultrasensitive detection and characterization of biomolecules using
  superchiral fields},\ }\href {https://doi.org/10.1038/nnano.2010.209}
  {\bibfield  {journal} {\bibinfo  {journal} {Nature Nanotechnology}\ }\textbf
  {\bibinfo {volume} {5}},\ \bibinfo {pages} {783} (\bibinfo {year}
  {2010})}\BibitemShut {NoStop}%
\end{thebibliography}%

\end{document}